\documentclass[lettersize,journal]{IEEEtran}
\usepackage{amssymb}
\usepackage{amsmath}
\usepackage{cite}
\usepackage{url}
\usepackage{xcolor}
\usepackage{cite,graphicx,amsmath,amssymb}
\usepackage{subfigure}
\usepackage{citesort}
\usepackage{fancyhdr}
\usepackage{mdwmath}
\usepackage{mdwtab}
\usepackage{caption}
\usepackage{amsthm}
\usepackage{setspace}
\usepackage{algorithm}
\usepackage{algorithmic}
\usepackage{makecell}
\usepackage{diagbox}
\usepackage{stfloats}
\usepackage{balance} 
\newcommand{\bm}[1]{\mbox{\boldmath{$#1$}}}

\newtheorem{theorem}{Theorem}

\newtheorem{lemma}{Lemma}

\newtheorem{corollary}{Corollary}

\allowdisplaybreaks
\makeatletter
\def\ScaleIfNeeded{
\ifdim\Gin@nat@width>\linewidth \linewidth \else \Gin@nat@width
\fi } \makeatother
 
\begin{document}
\title{STAR-RIS Assisted Downlink Active and Uplink Backscatter Communications with NOMA}
\author{
	Ao~Huang,
	Xidong~Mu,~\IEEEmembership{Member,~IEEE,}
	and
	Li~Guo,~\IEEEmembership{Member,~IEEE}
    \thanks{Ao Huang and Li Guo are with the Key Laboratory of Universal Wireless Communications, Ministry of Education, Beijing University of Posts and Telecommunications, Beijing 100876, China, also with the School of Artificial Intelligence, Beijing University of Posts and Telecommunications, Beijing 100876, China, and also with the National Engineering Research Center for Mobile Internet Security Technology, Beijing University of Posts and Telecommunications, Beijing 100876, China (email: huangao@bupt.edu.cn; guoli@bupt.edu.cn).}
	\thanks{Xidong Mu is with the School of Electronic Engineering and Computer Science, Queen Mary University of London, London E1 4NS, U.K. (e-mail:xidong.mu@qmul.ac.uk).}
}

\maketitle
\vspace{-1.5cm}
\begin{abstract}
A simultaneously transmitting and reflecting reconfigurable intelligent surface (STAR-RIS) assisted downlink (DL) active and uplink (UL) backscatter communication (BackCom) framework is proposed. More particularly, a full-duplex (FD) base station (BS) communicates with the DL users via the STAR-RIS's transmission link, while exciting and receiving the information from the UL BackCom devices with the aid of the STAR-RIS's reflection link. Non-orthogonal multiple access (NOMA) is exploited in both DL and UL communications for improving the spectrum efficiency. The system weighted sum rate maximization problem is formulated for jointly optimizing the FD BS active receive and transmit beamforming, the STAR-RIS passive beamforming, and the DL NOMA decoding orders, subject to the DL user's individual rate constraint. To tackle this challenging non-convex problem, we propose an alternating optimization (AO) based algorithm for the joint active and passive beamforming design with a given DL NOMA decoding order. To address the potential high computational complexity required for exhaustive searching all the NOMA decoding orders, an efficient NOMA user ordering scheme is further developed. Finally, numerical results demonstrate that: i) compared with the baseline schemes employing conventional RISs or space division multiple access, the proposed scheme achieves higher performance gains; and ii) higher UL rate gain is obtained at a cost of DL performance degradation, as a remedy, a more flexible performance tradeoff can be achieved by introducing the STAR-RIS.
\end{abstract}
\section{Introduction}
The Internet of Things (IoT) promises to enable the networked interconnection of dedicated device endpoints, extending a wide range of enhanced services based on new information technologies, from personal smart healthcare to smart city applications~\cite{al2015internet,dhillon2017wide}. 
In the upcoming sixth generation (6G) wireless communications, one of the major fundamental issues is to  ensure quality of service (QoS) for IoT devices, which are required to be connected in a spectrum- and energy-efficient manner. To this end, hybrid active and passive radio transmission is emerging as an innovative wireless paradigm for future communications~\cite{long2019symbiotic}. To support passive IoT, a great deal of existing works have been devoted to the smart backscatter communication (BackCom) technique, which has made a unique contribution to alleviate the energy issues caused by the access of large-scale wireless devices~\cite{boyer2014backscatter,xu2018practical}. Specifically, rather than generating its own radio-frequency (RF) signal to permit information exchange, an IoT device equipped with BackCom circuits can be excited by the received continuous wave (CW) signals and then modulate the message onto the incident signals for passive transmission. Since there is no need of power-hungry RF chains, BackCom consumes significantly less power compared to active radio transmission. 

Another promising solution to sustain IoT networks with low-power consumption relies on the development of reconfigurable intelligent surfaces (RISs)~\cite{saad2020vision}. Each RIS incorporates a large number of reflection units with tunable electromagnetic properties, which can actively regulate the space electromagnetic waves~\cite{wu2019towards}. However, the deployment of conventional RISs in a wireless environment can only benefit the devices located on one side by applying the smart control to the incident signals. To overcome this fundamental limitation, the simultaneously transmitting and reflecting reconfigurable intelligent surface (STAR-RIS) was proposed~\cite{liu2021STAR}. By supporting both electric-polarization and magnetization currents, each element on STAR-RIS can simultaneously reconfigure the transmitted and reflected signals, and thus achieving a highly flexible \textit{full-space} coverage for IoT applications.
\subsection{Prior Works}
\subsubsection{Studies on BackCom Systems}
A well-known application of BackCom is RF identification, where the tag simply powered by a high power reader to passively feedback information~\cite{boyer2014backscatter,galappaththigelink}. In the ambient BackCom (AmBC)~\cite{Van2018ambient}, direct communication between two backscatter transceivers can be supported via surrounding ambient RF sources, e.g., cellular base stations (BSs), Wi-Fi devices, and TV towers. More recently, research on BackCom has expanded to include a variety of scenarios. For instance, the authors of~\cite{li2019capacity} characterized the capacity performance of a fundamental BackCom system, in which the performance gain was improved by tag selection. The authors of~\cite{lu2018wireless} introduced a hybrid model combining ambient backscattering with wireless-powered communications, on the basis of which a novel device-to-device (D2D) communication paradigm was studied. The authors of~\cite{lyu2018throughput} considered a novel cognitive wireless powered communication network design, where two working modes were proposed for backscattering information in the secondary transmission system. Moreover, the authors of~\cite{ye2020outage} have investigated the outage performance for an ambient BackCom system with multiple backscatter links. Specifically, in order to minimize the outage probability of the selected backscatter link, an adaptive reflection coefficient strategy was conceived. To further improve the system efficiency, some prior works have integrated BackCom with RISs. Particularly, the authors of~\cite{jiaintelligent} made a pioneering study of the RIS assisted bistatic backscatter networks, aiming to minimize the transmit power at the carrier emitter based on the joint design of transmit beamforming and RIS phase-shift coefficients. The authors of~\cite{yanganalytical} combine the advantages of RIS and AmBC transmission, where the D2D terminals are able to flexibly select the operating mode to construct better channel conditions. The authors of~\cite{galappaththigeris} demonstrated that deploying RISs between the tag and the reader provides significant performance gains in a RIS-empowered AmBC system. 
\subsubsection{Studies on STAR-RIS assisted Systems}
Recently, research on the STAR-RIS assisted communications has been in full swing. The authors of ~\cite{xu2021star} initially reported the concept of STAR-RIS and analyzed the generic hardware design. The authors of~\cite{mu2021simultaneously} proposed three operating protocols for facilitating transmission by employing STAR-RISs, where the total transmit power was minimized by a joint beamforming design. In~\cite{wang2022coupled}, the phase-shifts of the STAR-RIS elements were assumed to be coupled, under this assumption, a more general optimization framework was proposed. The authors of~\cite{zhang2023security} further investigated the secure communication problem for coupled phase-shift STAR-RIS networks, in which the fair secrecy requirement for each user was guaranteed. Moreover, the authors of~\cite{wu2021coverage} presented the coverage characterization based on a proposed STAR-RIS aided transmisison framework. To achieve high spectral efficiency, the authors of~\cite{liu2022effective} proposed an STAR-RIS assisted transmission model based on non-orthogonal multiple access (NOMA) strategy, and the analysis of system effective capacity was presented. In another aspect, the authors of~\cite{li2022enhancing} have explored the secrecy performance for the STAR-RIS assisted NOMA network taking into account the residual hardware impairments.
\subsection{Motivations and Contributions}
Despite the burgeoning research, establishing large-scale regional IoT connectivity on a global scale is still in its infancy. One of the dilemmas we need to overcome is the power supply for IoT devices, especially when it comes to connecting a large number of devices. In this context, the hybrid network with coexistence of passive and active transmissions constitutes a green communication paradigm for future IoT. By encouraging simultaneous energy and spectrum cooperation between the different modes of transmissions, potential high performance gains can be introduced. The motivation of this paper stems from the observation it is appealing to apply BackCom technology for constructing hybrid radio networks, in which active RF sources can be utilized to simultaneously sustain passive transmission. Nonetheless, there exist a fundamental limitation to BackCom technology due to the ``double-channel fading'' effect~\cite{li2019capacity}, hindering its passive transmission for high data rate. On the other hand, long-distance communication can also lead to a lack of reliability for active transmission, especially in outdoor environments where the wireless links may be blocked by obstacles. Bearing all this in mind, in this paper, we focus our attention on the application of STAR-RISs in hybrid radio systems to achieve desired \textit{full}-space coverage enhancement, which is envisioned to open an entirely new realm for IoT.

In the meantime, the increasing number of IoT devices connected over a given frequency resource also poses a challenge for radio resource management.
To circumvent this issue, NOMA is a promising technique to facilitate massive connectivity~\cite{liu2022evolution}. Distinguished from conventional orthogonal multiple access (OMA) strategies, NOMA supports ubiquitous connectivity more efficiently by accommodating multiple users to occupy the same orthogonal resource block (RB). The prominent features of resource sharing make NOMA a powerful candidate for multiple access, providing flexible resource management for both active and passive transmissions in large-scale IoT networks.

Given these motivations, we propose a novel STAR-RIS assisted hybrid transmission framework, where the uplink (UL) passive BackCom is parasitic in the downlink (DL) active RF communication with the aid of STAR-RIS. Specifically, one full-duplex (FD) BS is designed to send information to the DL users and activate the UL BackCom users simultaneously. Moreover, NOMA protocol is applied in both active and passive communications to mitigate the potential stringent inter-user interference. The main contributions of this paper can be summarized as follows:
\begin{itemize}
	\item 
	We propose a STAR-RIS assisted DL active and UL BackCom framework, where a FD BS exchanges information with multiple DL and UL users simultaneously with the aid of NOMA and an STAR-RIS. Based on the proposed transmission framework, we formulate a weighted sum rate maximization problem by jointly designing the FD BS active beamforming vectors, the STAR-RIS passive beamforming coefficients and the DL NOMA decoding order, while guaranteeing the target rate constraint for each individual DL user.
	\item 
	We propose an alternating optimization (AO) based algorithm to cope with the intrinsically coupled non-convex problem, where the FD BS transmit and receive beamforming coefficients, and the passive STAR-RIS transmission and reflection beamforming coefficients are optimized in an alternating manner. In particular, the optimal receive beforming vector at the FD BS is derived in closed-form and a penalty based approach is proposed to address the non-convexity of rank-one constraint in STAR-RIS beamforing design. Given the fact that exhaustive search method requires a relatively high computational complexity, we conceive a combined channel gain based scheme to determine the DL NOMA decoding order, which reaps a good trade-off between computational complexity and optimality.
	\item 
	Our numerical results unveil that the proposed STAR-RIS assisted DL active and UL BackCom framework significantly outperforms systems employing conventional RISs or using space division multiple access (SDMA) and zero-forcing (ZF) detection for UL transmission.
	In hybrid transmssion networks, exploiting RF sources to simultaneously sustain active and passive communications yields an interesting performance tradeoff between the two transmission modes. In particular, by utilizing the unique degree-of-freedom (DoF) provided by the STAR-RIS, a more flexible tradeoff can be achieved.
\end{itemize}
\vspace{-0.5cm}
\subsection{Organization and Notation}
The rest of this paper is organized as follows. Section II presents the system model, based on which the optimization problem for designing the STAR-RIS assisted DL active and UL BackCom network is formulated. In Section III, an AO based algorithm is developed for solving the resulting non-convex problem, where a penalty based approach is conceived to achieve rank-one constraint relaxation.
Section IV proposes a low-complexity DL user ordering scheme. Simulation results are presented in Section V, which is followed by conclusions in Section VI.\\
\indent \emph{Notations:} Scalars, vectors, and matrices are denoted by lower-case, bold-face lower-case, and bold-face upper-case letters, respectively. ${\mathbb{C}^{N \times 1}}$ denotes the space of $N \times 1$ complex-valued vectors. ${{\mathbf{a}}^H}$ and $\left\| {\mathbf{a}} \right\|$ denote the conjugate transpose and the Euclidean norm of vector ${\mathbf{a}}$, respectively. ${\textup {diag}}\left( \mathbf{a} \right)$ denotes a diagonal matrix with the elements of vector ${\mathbf{a}}$ on the main diagonal. The distribution of a circularly symmetric complex Gaussian (CSCG) random variable with mean $\mu $ and variance ${\sigma ^2}$ is denoted by ${\mathcal{CN}}\left( {\mu,\sigma _k^2} \right)$. ${{\mathbf{1}}_{m \times n}}$ and ${{\mathbf{0}}_{m \times n}}$ denote the all-one and an all-zero matrices of size ${m \times n}$, respectively. ${\mathbb{H}^{N}}$ denotes the set of all $N$-dimensional complex Hermitian matrices. ${\textup {Rank}}\left( \mathbf{A} \right)$ and ${\textup {Tr}}\left( \mathbf{A} \right)$ denote the rank and the trace of matrix $\mathbf{A}$, respectively. ${\textup {diag}}\left( \mathbf{A} \right)$ denotes a vector whose elements are extracted from the main diagonal elements of matrix $\mathbf{A}$. ${{\mathbf{A}}} \succeq 0$ indicates that $\mathbf{A}$ is a positive semidefinite matrix. ${\left\| {\mathbf{A}} \right\|_*}$, ${\left\| {\mathbf{A}} \right\|_2}$, and ${\left\| {\mathbf{A}} \right\|_F}$ denote the nuclear norm, spectral norm, and Frobenius norm of matrix $\mathbf{A}$, respectively.
\section{System Model and Problem Formulation}
\subsection{System Model }
\begin{figure}[t]
	\centering
	\setlength{\belowcaptionskip}{+0.2cm} 
    \includegraphics[width=3.5in]{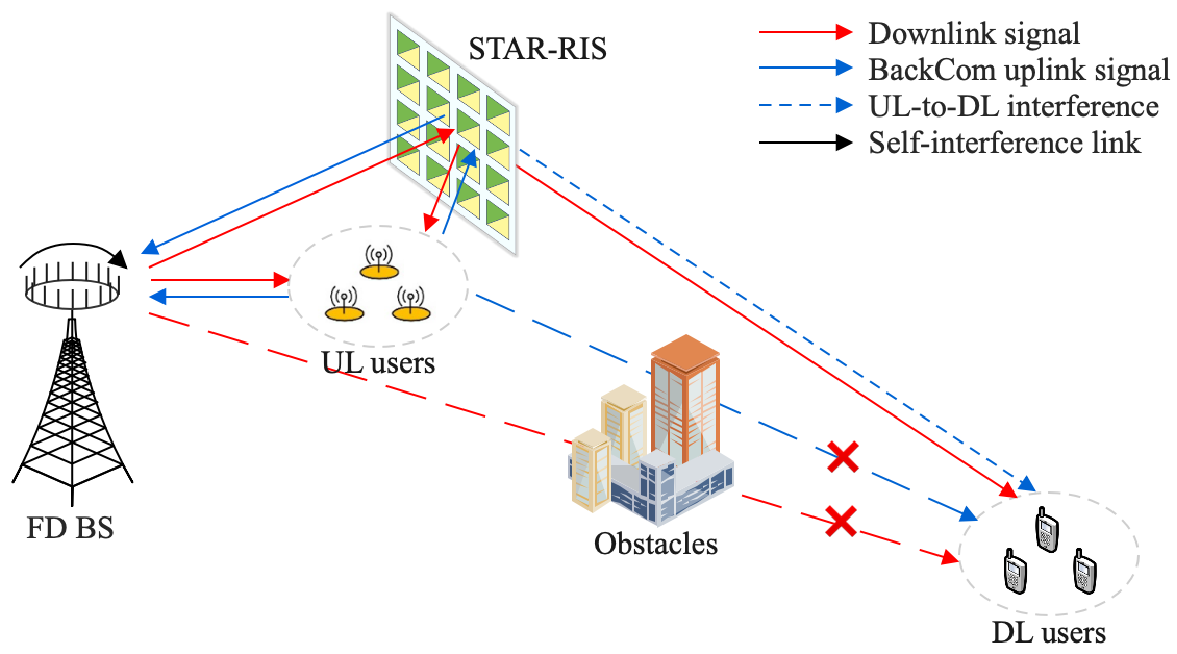}
	\caption{Illustration of the STAR-RIS assisted DL active and UL backscatter communications. }
	\label{System model}
\end{figure}
As illustrated in Fig. 1, we consider a STAR-RIS assisted DL active and UL passive BackCom framework, where the FD BS is equipped with $N$ antennas and the STAR-RIS consists of $M$ elements denoted by the set $\mathcal{M}=\{1,2,\cdots,M\}$.
Simultaneous UL reception and DL transmission of the FD BS can be carried out within the same frequency band resource~\cite{bharadia2013full}. There are $K$ UL BackCom users and  $J$ DL users whose users are indexed by the sets $\mathcal{K}=\{1, 2,\cdots, K\}$ and $\mathcal{J}=\{1, 2,\cdots, J\}$, respectively. All DL and UL users are equipped with one single antenna. An STAR-RIS operating in the energy splitting (ES) mode~\cite{mu2021simultaneously} is deployed in the vicinity of the FD BS. Each element on STAR-RIS can actively adjust the amplitudes and phase-shifts of the incident signals to realize controllable signal transmission and reflection. Let $\mathbf{ \Theta}_t\in\mathbb{C}^{M\times M}$ and $\mathbf{ \Theta}_r\in\mathbb{C}^{M\times M}$ denote transmission- and reflection-coefficient matrices, respectively, which can be modeled as
\begin{equation}
	\begin{aligned}
	&\mathbf{ \Theta}_\chi= {\textup{diag}}\{\sqrt{\beta_1^\chi }e^{j\theta_{1}^\chi }, \sqrt{\beta_2^\chi }e^{j\theta_{2}^\chi },\cdots, \sqrt{\beta_M^\chi }e^{j\theta_{M}^\chi }\}, \\
	&\qquad\qquad\qquad\qquad\qquad\qquad\qquad\qquad\forall \chi \in\{t,r\}.
	\end{aligned}
\end{equation}
In the above expression, we have $\beta_m^t+\beta_m^r=1$ due to the law of energy conservation, $\theta_{m}^t, \theta_{m}^r \in [0,2\pi), \forall m\in\mathcal{M}$ denote the reflection and transmission phase-shift coefficients of the $m$-th element, which is assumed to be continuously adjustable to investigate the maximum system performance$\footnote{In this paper, we assume that the phase-shift coefficients for transmission and reflection can be adjusted independently. Note that with appropriate modifications, the developed algorithms can be generalized to the case that the STAR-RIS adopts a coupled phase-shift model~\cite{wang2022coupled}.}$. The whole space is divided into two half-spaces by the STAR-RIS. Without loss of generality, we assume that the UL BackCom users are located in the reflection region of the STAR-RIS and can exchange information directly with the FD BS via a direct link in addition to the help of the STAR-RIS. The DL users are located in the transmission region of the STAR-RIS and can only receive signals transmitted by the STAR-RIS.
In practice, these UL BackCom users could be a number of sensors to upload the information containing in a few bytes. Whereas, the DL users could be some terminals, i.e., mobile phones, etc., which can be served with higher target data rate$\footnote{For more general communication scenario settings, such as different types of devices randomly distributed on both sides of the STAR-RIS, this requires more sophisticated transmission mechanisms and is therefore left as our future work.}$.

Let $\mathbf{g}_k\in\mathbb{C}^{1\times N}$ and $\mathbf{g}_k^H\in\mathbb{C}^{N\times 1}$ denote the channels between the FD BS and the UL user $k$. $\mathbf{f}_{\rm{BS}}\in\mathbb{C}^{M \times N}$ and $\mathbf{f}_{\rm{SB}}^H\in\mathbb{C}^{N \times M}$ denote the channels between the FD BS and the STAR-RIS. $\mathbf{h}_{k,{\rm S}}\in\mathbb{C}^{M \times 1}$ and $\mathbf{h}_{{\rm S},k}^H\in\mathbb{C}^{1 \times M}$ denote the channels between the STAR-RIS and UL BackCom user $k$. $\mathbf{h}_{j,{\rm S}}\in\mathbb{C}^{M \times 1}$ and $\mathbf{h}_{{\rm S},j}^H\in\mathbb{C}^{1 \times M}$ denote the channels between the STAR-RIS and DL user $j$. It is assumed that the channel state information (CSI) of the proposed wireless communication system has been obtained by the advanced channel estimation methods~\cite{wu2021channel,ding2021On}. 
\subsection{NOMA Based UL and DL Communications}
NOMA is employed in both UL and DL communications for improving system spectrum efficiency, the transmit signal vector generated by the FD BS is given by
\begin{equation}
	\mathbf{x}_0=\sum\limits_{j\in\mathcal{J}}\mathbf{w}_j s_j,
\end{equation}
where $\mathbf{w}_j\in\mathbb{C}^{N \times 1}$ denotes the active transmit beamforming vector for DL user $j$. $s_j$ denotes the corresponding information-bearing
symbol with $\mathbb{E}\{\left|s_j\right|^2\}= 1, \forall j\in\mathcal{J}$.
Regarding the power requirements at the FD BS, we have the following constraint,
\begin{equation}
	\mathbb{E}\{\left\|\mathbf{x}_0\right\|^2\}=\sum\limits_{j\in\mathcal{J}}\left\|\mathbf{w}_j\right\|^2\le P_{\max},
\end{equation}
where $P_{\max}$ denotes the maximum power allowance at the FD BS to send messages.
\subsubsection{UL passive backscatter communication}
The communication between the BS and the UL BackCom users can be enhanced via the reflection link of the STAR-RIS. Given the transmit signal at the FD BS, the signal received at UL user $k$ is given by 
\begin{equation}
	y_{k}^{\rm UL}=\left(\mathbf{g}_k+\mathbf{h}_{{\rm S},k}^H\mathbf \Theta_r\mathbf{f}_{\rm BS}\right)\mathbf{x}_0, \forall k\in\mathcal{K}.
\end{equation}
 Each UL user then facilitates the exchange of information by passively transmitting signals to the FD BS. Let  $c_k$ denote the UL user's own signal with $\mathbb{E}\{\left|c_k\right|^2\}= 1$. The BackCom reflection coefficient $\eta_k$ at user $k$ satisfies $0\le\eta_k\le1$. 
We assume $\eta_k = 1,  \forall k\in\mathcal{K}$, indicating that all the power of the RF signals is captured by UL users to permit backscattering modulation$\footnote{{In this paper, we assume that data transmission causes the main energy consumption of IoT users, and for simplicity the hardware consumption of the circuit is not considered~\cite{ju2013throughput}.}}$. The backscattered signal at UL user $k$ can be given by $x_{k}=\left(\mathbf{g}_k+\mathbf{h}_{{\rm S},k}^H\mathbf \Theta_r\mathbf{f}_{\rm BS}\right)\mathbf{x}_0c_k$. Assuming that perfect synchronization is performed at the FD BS, the received signal $\mathbf{y}_{\rm B}$ can be written as
\begin{equation}
	\mathbf{y}_{\rm B}=\sum\limits_{k\in\mathcal{K}}\left(\mathbf{g}_k^H+\mathbf{f}_{\rm SB}^H\mathbf \Theta_r\mathbf{h}_{k,{\rm S}}\right)x_k+\mathbf{H}_{\rm SI}\mathbf{x}_0+\mathbf{n}_{\rm B},
\end{equation}
where  $\mathbf{H}_{\rm SI}\in\mathbb{C}^{N \times N}$ denotes the self-interference (SI) channel gain at the FD BS. $\mathbf{n}_{\rm B}\sim \mathcal{CN}\left(0, \sigma_{\rm B}^2\mathbf{I}_{N}\right)$ is the AWGN with noise power $\sigma_{\rm B}^2$.

Based on the UL NOMA principle, inter-user interference due to resource sharing can be properly overcome by implementing successive interference cancellation (SIC) at the FD BS. A receive beamforming vector $\mathbf{z}\in\mathbb{C}^{N \times 1}$ is firstly exploited at the multi-antenna BS to linearly detect the the superimposed signal $\mathbf{y}_{\rm B}$~\cite{havary2010joint}. As a result, the extracted signal through the receive beamforming $\mathbf{z}$ is given by
\begin{equation}
	\begin{aligned}
	\widetilde{y}_{\rm B}&=\mathbf{z}^H\mathbf{y}_{\rm B}\\
	&=\underbrace{\mathbf{z}^H\sum\limits_{k\in\mathcal{K}}\widetilde{\mathbf{h}}_k^H\widetilde{\mathbf{h}}_k\mathbf{x}_0c_k}_{\textup{desired signal}}+\underbrace{\mathbf{z}^H\mathbf{H}_{\rm SI}\mathbf{x}_0+\mathbf{z}^H\mathbf{n}_{\rm B}}_{\textup{aggregated noise}},
		\end{aligned}
\end{equation}
where $\widetilde{\mathbf{h}}_k=\mathbf{g}_k+\mathbf{h}_{{\rm S},k}^H\mathbf \Theta_r\mathbf{f}_{\rm BS}\in\mathbb{C}^{1 \times N}$ and $\widetilde{\mathbf{h}}_k^H=\mathbf{g}_k^H+\mathbf{f}_{\rm SB}^H\mathbf \Theta_r\mathbf{h}_{k,{\rm S}}\in\mathbb{C}^{N \times 1}$ denote the effective channels between the FD BS and UL user $k$. 
Before decoding the UL users’ backscattered signals, we first apply SI cancellation techniques to significantly reduce the power gain of SI signal~\cite{elhattab2021reconfigurable}. Note that the decoding order of the UL NOMA does not affect the achievable UL sum rate~\cite{diamantoulakis2016wireless}, which is given by
\begin{equation}
	R_{sum}^{\rm UL}=\log_2\left(1+\frac{\sum\limits_{k\in\mathcal{K}}\left|\mathbf{z}^H\tilde{\mathbf{h}}_k^H\tilde{\mathbf{h}}_k\mathbf{x}_0\right|^2}{\rho\left|\mathbf{z}^H\mathbf{H}_{\rm SI}\mathbf{x}_0\right|^2+\sigma_{\rm B}^2\left\|\mathbf{z}\right\|^2}\right),
\end{equation}
where $0\le\rho\ll1$ indicates the amount of residual SI at the FD BS.
\subsubsection{DL active communication}
The FD BS sends messages to the DL users via the transmission link of the STAR-RIS. Meanwhile, the DL users will also be interfered by the backscattered signals from the UL BackCom users. Therefore, the signal received at DL user $j$ is given by
\begin{equation}
	\begin{aligned}
	y_{j}^{\rm DL}=&\left(\mathbf{h}_{{\rm S},j}^H\mathbf \Theta_t\mathbf{f}_{\rm BS}\right)\mathbf{w}_js_j+\underbrace{{\sum_{i\in\mathcal{J}\backslash \{j\}}\left(\mathbf{h}_{{\rm S},j}^H\mathbf \Theta_t\mathbf{f}_{\rm BS}\right)\mathbf{w}_is_i}}_{\textup{multiuser interference}}\\
	&+\underbrace{\sum\limits_{k\in\mathcal{K}}\left(\mathbf{h}_{{\rm S},j}^H\mathbf \Theta_t\mathbf{h}_{k,{\rm S}}\right)x_k}_{\textup{UL-to-DL interference}}+n_{j}, \forall j\in\mathcal{J},
		\end{aligned}
\end{equation}
where $n_{j}\sim \mathcal{CN} (0, \sigma^2)$ denotes the AWGN at DL user $j$ with noise power $\sigma^2$. Compared with conventional DL transmission system, each DL user suffers from performance degradation due to the UL-to-DL co-channel interference, i.e., the term $I_j\triangleq\sum_{k\in\mathcal{K}}\left(\mathbf{h}_{{\rm S},j}^H\mathbf \Theta_t\mathbf{h}_{k,{\rm S}}\right)x_k$ in (8), whose power is given by $\mathcal{E}\left\{I_jI_j^*\right\}=\sum_{k\in\mathcal{K}}\left|\left(\mathbf{h}_{{\rm S},j}^H\mathbf \Theta_t\mathbf{h}_{k,{\rm S}}\right)\tilde{\mathbf{h}}_k\mathbf{x}_0\right|^2$.

Following the DL NOMA principle, each user employs the SIC technique to remove the multiuser interference.
More particularly, the user with a stronger combined channel gain can first decode the signals of the users with the weaker combined channel gains, before decoding its own signal.
Let $\Omega_j\in\mathcal{J}$ denote the decoding order for DL user $j$. For a given order, the following constraints should be satisfied to accommodate the rate fairness conditions,
\begin{equation}
	\begin{aligned}
	{\left|\left(\mathbf{h}_{{\rm S},i}^H\mathbf \Theta_t\mathbf{f}_{\rm BS} \right)\mathbf{w}_j\right|^2}\ge{\left|\left(\mathbf{h}_{{\rm S},i}^H\mathbf \Theta_t\mathbf{f}_{\rm BS}\right)\mathbf{w}_l\right|^2},\;&{\rm if}\;\Omega_j<\Omega_l,\\
	&\forall i,j,l\in\mathcal{J}.
		\end{aligned}
\end{equation}
The inequalities in (9) is imposed to ensure that more wireless resources are allocated to DL users with a lower decoding order~\cite{mu2020exploiting}.
Therefore, the achievable signal-to-interference-plus-noise ratio (SINR) at DL user $j$ to decode its own signal is given by
\begin{equation}
	{\rm SINR}_{j\rightarrow j}=\frac{{\left|\left(\mathbf{h}_{{\rm S},j}^H\mathbf \Theta_t\mathbf{f}_{\rm BS}\right)\mathbf{w}_j\right|^2}}{\mathcal{E}\{I_jI_j^*\}+{\sum\limits_{\Omega_i>\Omega_j}\left|\left(\mathbf{h}_{{\rm S},j}^H\mathbf \Theta_t\mathbf{f}_{\rm BS}\right)\mathbf{w}_i\right|^2}+\sigma^2}.
\end{equation}
The corresponding achievable rate at DL user $j$ to decode its own signal is $R_{j\rightarrow j}=\log_2\left(1+{\rm SINR}_{j\rightarrow j}\right)$. 
Furthermore, for any DL user $j,l\in\mathcal{J}$ which satisfy $\Omega_j<\Omega_l$, the SINR at user $l$ to decode user $j$’s signal is given by
\begin{equation}
	{\rm SINR}_{l\rightarrow j}=\frac{{\left|\left(\mathbf{h}_{{\rm S},l}^H\mathbf \Theta_t\mathbf{f}_{\rm BS}\right)\mathbf{w}_j\right|^2}}{\mathcal{E}\{I_lI_l^*\}+{\sum\limits_{\Omega_i>\Omega_j}\left|\left(\mathbf{h}_{{\rm S},l}^H\mathbf \Theta_t\mathbf{f}_{\rm BS}\right)\mathbf{w}_i\right|^2}+\sigma^2}.
\end{equation}
The corresponding achievable rate at DL user $l$ to decode user $j$'s signal is $R_{l\rightarrow j}=\log_2\left(1+{\rm SINR}_{l\rightarrow j}\right)$. 

Denote $R_{j}^{\rm DL}$ as the data rate of DL user $j$ to decode its own signal. 
However, this rate is achievable if and only if the stronger DL user can successfully decode the signals of the weaker users~\cite{mu2020exploiting}. To guarantee effective implementation of the SIC, the following rate constraints are imposed:
\begin{equation}
	\min\left\{R_{j\rightarrow j}, R_{l\rightarrow j}\right\}\ge R_{j}^{\rm DL},  \;{\rm if}\;\Omega_j<\Omega_l, \forall j,l\in\mathcal{J}.
\end{equation}
In a nutshell, the achievable sum rate of all $J$ DL users is  ${R_{sum}^{\rm DL}}=\sum_{j\in\mathcal{J}}R_{j}^{\rm DL}$.
\subsection{Problem Formulation }
In this paper, our objective is to maximize the system weighted sum rate based on the joint design of the FD BS active receive and transmit beamforming, the STAR-RIS passive beamforming and the NOMA decoding orders, subject to the DL user's individual rate constraint. The optimization problem can be formulated as follows:
\begin{subequations}\label{P1}
	\begin{align}
    	&\mathop{\rm{max}}\limits_{\mathbf{z},\{\mathbf{w}_j,\mathbf{ \Theta}_{\chi },\Omega_j\}}\quad \omega^{\rm UL}{R_{sum}^{\rm UL}}+\omega^{\rm DL}{R_{sum}^{\rm DL}}\\
		&\label{P1_C1}{\rm s.t.} \;
		\sum_{j\in\mathcal{J}}\left\|\mathbf{w}_j\right\|^2\le P_{\max},\\
		&\label{P1_C2}\quad
		{\left|(\mathbf{h}_{{\rm S},i}^H\mathbf \Theta_t\mathbf{f}_{\rm BS})\mathbf{w}_j\right|^2}\ge{\left|(\mathbf{h}_{{\rm S},i}^H\mathbf \Theta_t\mathbf{f}_{\rm BS})\mathbf{w}_l\right|^2}, \notag\\
		&\quad
		{\rm if}\;\Omega_j<\Omega_l,
		\forall i,j,l\in\mathcal{J},\\
		&\label{P1_C3}\quad
		\min\left\{R_{j\rightarrow j}, R_{l\rightarrow j}\right\}\ge R_{j}^{\rm DL},  \;{\rm if}\;\Omega_j<\Omega_l, \forall j,l\in\mathcal{J},\\
		&\label{P1_C4}\quad
	    R_{j}^{\rm DL}\ge \overline{R},\forall j\in\mathcal{J},\\
		&\label{P1_C5}\quad
		\theta_m^t, \theta_m^r\in[0,2\pi), \forall m\in\mathcal{M},\\
		&\label{P1_C6}\quad
		\beta_m^t,\beta_m^r\in[0,1], \beta_m^t+\beta_m^r=1,\forall m\in\mathcal{M},	    
	\end{align}
\end{subequations}
where $\omega^{\rm UL}\ge0$ and $\omega^{\rm DL}\ge0$ denote the predefined weights, which can be used to prioritize the UL and DL communications. Constraint \eqref{P1_C1} restricts the transmit power at the FD BS. 
Constraints \eqref{P1_C2} and \eqref{P1_C3} provide the conditions for successfully decoding the user signals using SIC.
Constraint \eqref{P1_C4} ensures the achievable rate of each DL user satisfies the target rate requirement $\overline R$. 
Constraints \eqref{P1_C5} and \eqref{P1_C6} represent the unimodular phase-shift constraint and the energy conservation constraint for each STAR-RIS element, respectively.
\subsection{Discussion}
Note that optimization problem \eqref{P1}  is non-convex, which cannot be handled directly. We summarize the main challenges for solving this problem as follows:  
1) the hybrid resource allocation coefficients (i.e., $\mathbf{z}$, $\mathbf{w}_j$, $\mathbf{ \Theta}_{\chi}$, and $\Omega_j$) are highly coupled, especially the unit-modulus constraint \eqref{P1_C5} for element phase-shift design on the STAR-RIS is non-convex; 
and 2) due to the parallelism of DL active transmission with UL passive BackCom, UL-to-DL interference needs to be taken into account when calculating DL data rate, which imposes additional complexity for optimization.

To the best of our knowledge, the globally optimal solution of the original problem is generally intractable.
In the following sections, we first decompose problem \eqref{P1} into several subproblems, which can be solved alternately with a given DL decoding order by invoking AO method. Then, a low-complexity NOMA decoding ordering strategy is further proposed for DL active communication. 
\section{Proposed Solution Based on Alternating Optimization}
In this section, based on the proposed transmission design, we focus on improving the system weighted sum rate for a given decoding order. In particular, the original problem \eqref{P1} is decomposed into three subproblems, i.e., active FD BS transmit beamforming optimization, active FD BS receive beamforming optimization, and joint passive transmission- and reflection- beamforming optimization at the STAR-RIS. 
\subsection{Active FD BS Transmit Beamforming Optimization}
For a given decoding order $\{\Omega_j\}$, we first optimize the active transmit beamforming $\{\mathbf{w}_j\}$ at the FD BS with fixed $\{\mathbf{z}\}$ and $\{\mathbf{ \Theta}_t,\mathbf{ \Theta}_r\}$. Problem \eqref{P1} is simplified as
\begin{subequations}\label{P2}
	\begin{align}	
		&\mathop{\rm{max}}\limits_{\{\mathbf{w}_j\}}\quad  \omega^{\rm UL}{R_{sum}^{\rm UL}}+\omega^{\rm DL}{R_{sum}^{\rm DL}}\\
		&\;\;{\rm s.t.} \quad
		\textup{\eqref{P1_C1} -- \eqref{P1_C4}}.
	\end{align}
\end{subequations}
The non-convexity of problem \eqref{P2} lies in the non-concave objective function and the non-convex  constraints \eqref{P1_C1} -- \eqref{P1_C3}. To address this issue, the FD BS transmit matrix is defined as $\mathbf{W}_j\triangleq{\mathbf{w}_j}{\mathbf{w}_j}^H\in\mathbb{C}^{N \times N}, \forall j\in\mathcal{J}$, which satisfies $\mathbf{W}_j\succeq 0$ and $\textup{Rank}(\mathbf{W}_j)=1$. Moreover, we denote ${\mathbf{H}}_k\triangleq\widetilde{\mathbf{h}}_k^H\widetilde{\mathbf{h}}_k$. After straightforward algebraic multiplications, the sum rate function of the UL NOMA system can be rewritten as
\begin{equation}
	R_{sum}^{\rm UL}=\log_2\left(1+\frac{\sum\limits_{k\in\mathcal{K}}\sum\limits_{j\in\mathcal{J}}\textup{Tr}\left(\mathbf W_j{\mathbf{H}}_k^H\mathbf{Z}{\mathbf{H}}_k \right)}{\sum\limits_{j\in\mathcal{J}}\rho\textup{Tr}\left(\mathbf W_j\mathbf{H}_{\rm SI}^H\mathbf{Z}\mathbf{H}_{\rm SI}\right)+\sigma_{\rm B}^2\textup{Tr}\left(\mathbf Z\right)}\right),
\end{equation}
where $\mathbf{Z}\triangleq{\mathbf{z}}{\mathbf{z}}^H\in\mathbb{C}^{N \times N}$ is the receive matrix at the FD BS.

Let $\widetilde{\mathbf{h}}_{k,j}\triangleq\mathbf{h}_{{\rm S},j}^H\mathbf \Theta_t\mathbf{h}_{k,{\rm S}}\in\mathbb{C}^{1 \times 1}$ and $\widetilde{\mathbf{h}}_{j}\triangleq\mathbf{h}_{{\rm S},j}^H\mathbf \Theta_t\mathbf{f}_{\rm BS}\in\mathbb{C}^{1 \times N}$ denote
the channel from UL user $k\in\mathcal{K}$ to DL user $j\in\mathcal{J}$ and the channel from the FD BS to DL user $j\in\mathcal{J}$, respectively.
Therefore, the power of the  UL-to-DL co-channel interference is given by
\begin{equation} \label{eqn2}
	\begin{split}
		\mathcal{E}\left\{I_jI_j^*\right\}=\sum\limits_{k\in\mathcal{K}}\left(\left|\widetilde {\mathbf{h}}_{k,j}\right|^2\left|\widetilde{\mathbf{h}}_k\widetilde{\mathbf{w}}\right|^2\right)
		=\sum\limits_{k\in\mathcal{K}}\left(p_k\left|\widetilde{\mathbf{h}}_{k,j}\right|^2\right),
	\end{split}
\end{equation}
where $\widetilde {\mathbf{w}}\triangleq[\mathbf{w}_1\cdots\mathbf{w}_J]\in\mathbb{C}^{N \times J}$ and $p_k\triangleq\sum_{j\in\mathcal{J}}\textup{Tr}\left(\mathbf W_j{\mathbf{H}}_k\right)$.
By introducing the slack variables $\{S_{lj}\}$ and $\{I_{lj}\}$ as follows
\begin{gather}
	\frac{1}{S_{lj}}=\left|\widetilde{\mathbf{h}}_{l}\mathbf{w}_j\right|^2=\textup{Tr}\left({\mathbf W}_j{\mathbf{H}}_l\right),\\
	I_{lj}=\sum\limits_{k\in\mathcal{K}}\left(p_k\left|\widetilde{\mathbf{h}}_{k,l}\right|^2\right)+\sum_{\Omega_i>\Omega_j}\textup{Tr}\left({\mathbf W}_i{\mathbf{H}}_l\right)+\sigma^2,
\end{gather}
where ${\mathbf{H}}_l\triangleq\widetilde{\mathbf h}_{l}^H\widetilde{\mathbf h}_{l}, \forall l\in\mathcal{J}$, we can rewrite the SINR term as 
\begin{equation}
	{\rm SINR}_{l\rightarrow j}=\frac{1}{S_{lj}I_{lj}}, {\Omega_l\ge\Omega_j}.
\end{equation}	

With the above variable definitions, problem \eqref{P2} is equivalent to
\begin{subequations}\label{P3}
	\begin{align}
		&\mathop{\rm{max}}\limits_{\{\mathbf{W}_j,R_{j}^{\rm DL},{S_{lj}},{I_{lj}}\}}\quad 
		 \omega^{\rm UL}{R_{sum}^{\rm UL}}+ \omega^{\rm DL}\sum\limits_{j\in\mathcal{J}}{R_{j}^{\rm DL}}\\
		&\label{P3_C1}{\rm s.t.} 
	    \sum_{j\in\mathcal{J}}\textup{Tr}\left({\mathbf W}_j\right)\le P_{\max},\\
		&\label{P3_C2}\quad
		\textup{Tr}\left({\mathbf W}_j{\mathbf{H}}_i\right)   \ge\textup{Tr}\left({\mathbf W}_l{\mathbf{H}}_i\right), \;{\rm if}\;\Omega_j<\Omega_l,
		\forall i,j,l\in\mathcal{J},\\
		&\label{P3_C3}\quad
		\frac{1}{S_{lj}}\le\textup{Tr}\left({\mathbf W}_j{\mathbf{H}}_l\right),	\forall j,l\in\mathcal{J},\\
		&\label{P3_C4}\quad
		I_{lj}\ge\sum\limits_{k\in\mathcal{K}}\left(p_k\left|\widetilde{\mathbf{h}}_{k,l}\right|^2\right)+\sum_{\Omega_i>\Omega_j}\textup{Tr}\left({\mathbf W}_i{\mathbf{H}}_l\right)+\sigma^2,\notag\\	
		&\quad\forall k\in\mathcal{K}, \forall j,l\in\mathcal{J},\\
		&\label{P3_C5}\quad
		\min\left\{\log_2\left(1+\frac{1}{S_{jj}I_{jj}}\right),\log_2\left(1+\frac{1}{S_{lj}I_{lj}}\right)\right\}\ge R_{j}^{\rm DL},\notag\\
		&\quad{\rm if}\;\Omega_j<\Omega_l,
		\forall j,l\in\mathcal{J},\\
		&\label{P3_C6}\quad
		\mathbf{W}_j \succeq 0,
		\textup{Rank}\left(\mathbf{W}_j \right)=1,\\
		& \quad
		\textup{\eqref{P1_C4}},
	\end{align}
\end{subequations}
where constraints \eqref{P3_C3} and \eqref{P3_C4} are imposed to ensure that constraint \eqref{P3_C5} is an equivalent translation of \eqref{P1_C3}. Due to the complication of logarithmic operations, it is difficult to directly convert the left-hand side (LHS) of \eqref{P3_C5} into the form of a standard concave function. We note, since its Hessian function is semidefinite for any ${S_{lj}}> 0$ and ${I_{lj}} > 0$, the logarithmic function is jointly convex with respect to ${S_{lj}}$ and ${I_{lj}}$. Recall that any convex function can be approximated by a first-order Taylor expansion at any point to characterize its global lower bound~\cite{boyd2004convex}, which facilitates the application of the successive convex approximation (SCA) technique~\cite{kurtaran2002crashworthiness}. Given local points $\{\widetilde S_{lj}, \widetilde I_{lj}\}$, we can construct the lower bound of $R_{jl}$ as follows,
\begin{equation}
	\begin{aligned}
		\log_2&\left(1+\frac{1}{S_{lj} I_{lj}}\right)\ge
		\left[R_{l\rightarrow j}\right]^{lb}\\
		\triangleq&\log_2\left(1+\frac{1}{\widetilde S_{lj} \widetilde I_{lj}}\right)-\frac{S_{lj}-\widetilde S_{lj}}{\ln2\left(\widetilde S_{lj}^2\widetilde I_{lj}+\widetilde S_{lj}\right)}\\
		&-\frac{I_{lj}-\widetilde I_{lj}}{\ln2\left(\widetilde I_{lj}^2\widetilde S_{lj}+\widetilde I_{lj}\right)}.
	\end{aligned}
\end{equation}
Due to (21), the non-convex constraint \eqref{P3_C5} can be relaxed to
\begin{equation}
	\min\left\{\left[R_{j\rightarrow j}\right]^{lb}, \left[R_{l\rightarrow j}\right]^{lb}\right\}\ge R_{j}^{\rm DL},  \;{\rm if}\;\Omega_j<\Omega_l, \forall j,l\in\mathcal{J}.
\end{equation}

As for the non-convex term ${R_{sum}^{\rm UL}}$ in the objective function (20a), we construct its global underestimator in a similar manner. By replacing $\left\{S_{lj}\right\}$ and $\left\{I_{lj}\right\}$ with $\tau$ and $\delta$, which satisfy
\begin{gather}
	\frac{1}{\tau}\le{\sum\limits_{k\in\mathcal{K}}\sum\limits_{j\in\mathcal{J}}\textup{Tr}\left(\mathbf W_j{\mathbf{H}}_k^H\mathbf{Z}{\mathbf{H}}_k \right)},\\
	\delta\ge{\sum\limits_{j\in\mathcal{J}}\rho\textup{Tr}\left(\mathbf W_j\mathbf{H}_{\rm SI}^H\mathbf{Z}\mathbf{H}_{\rm SI}\right)+\sigma_{\rm B}^2\textup{Tr}\left(\mathbf Z\right)},
\end{gather}
respectively, the UL sum rate function can be rewritten as
\begin{equation}
	\begin{aligned}
		\log_2&\left(1+\frac{1}{\tau \delta}\right)\ge
		\left[R_{sum}^{\rm UL}\right]^{lb}\\
		\triangleq&\log_2\left(1+\frac{1}{\widetilde \tau \widetilde \delta}\right)-\frac{\tau-\widetilde \tau}{\ln2\left(\widetilde \tau^2\widetilde \delta+\widetilde \tau\right)}-\frac{\delta-\widetilde \delta}{\ln2\left(\widetilde \delta^2\widetilde \tau+\widetilde \delta\right)},
	\end{aligned}
\end{equation}
where $\widetilde \tau$ and $\widetilde \delta$ denote the local points generated in the previous iteration.

Then, problem \eqref{P3} can be reformulated as
\begin{subequations}\label{P4}
	\begin{align}
		&\mathop{\rm{max}}\limits_{\{\mathbf{W}_j,R_{j}^{\rm DL},R_{sum}^{\rm UL},{S_{lj}},{I_{lj}}\},\tau,\delta}\;
		\omega^{\rm UL}R_{sum}^{\rm UL}+ \omega^{\rm DL}\sum\limits_{j\in\mathcal{J}}{R_{j}^{\rm DL}}\\
		&\qquad\;\;{\rm s.t.} \quad
		R_{sum}^{\rm UL}\le\left[R_{sum}^{\rm UL}\right]^{lb},\\
		&\quad\quad\quad\quad\quad\textup{\eqref{P1_C4}, \eqref{P3_C1}  -- \eqref{P3_C4}, \eqref{P3_C6}, (22) -- (24)}.
	\end{align}
\end{subequations}
According to~\cite{xu2020resource}, the solutions $\{\mathbf{W}_j\}$ obtained by ignoring $\textup{Rank}(\mathbf{W}_j )=1$ for active beamforming design in problem \eqref{P4} always satisfy the rank-one constraint. 
Note that the final relaxed problem without constraint \eqref{P3_C6} is a standard semidefinite programming (SDP)~\cite{luo2010semidefinite}, which can be successfully solved with the aid of the existing convex optimization solvers such as CVX~\cite{grant2014cvx}. After solving problem \eqref{P4}, we can recover the active transmit beamforming vector $\{\mathbf{w}_j\}$ for problem \eqref{P2} via Cholesky decomposition, i.e., $\mathbf{W}_j^*={\mathbf{w}_j}{\mathbf{w}_j}^H, j\in\mathcal{J}$, where $\mathbf{W}_j^*$ denotes the optimal solution to problem \eqref{P4}.
\subsection{Active FD BS Receive Beamforming Optimization}
Note that the optimization variable ${\mathbf{z}}$ is only appealed in the UL sum rate function. For given $\{\mathbf{w}_j\}$ and $\{\mathbf{ \Theta}_t,\mathbf{ \Theta}_r\}$, the subproblem for optimizing the active FD BS receive beamforming is reduced to
\begin{equation}\label{P5}
	\mathop{\rm{max}}\limits_{\mathbf{z}}\quad {R_{sum}^{\rm UL}}\triangleq\log_2\left(1+\Gamma_{\rm B}\right),
\end{equation}
where $\Gamma_{\rm B}=\frac{\sum_{k\in\mathcal{K}}\left|\mathbf{z}^H\widetilde{\mathbf{h}}_k^H\widetilde{\mathbf{h}}_k\mathbf{x}_0\right|^2}{\rho\left|\mathbf{z}^H\mathbf{H}_{\rm SI}\mathbf{x}_0\right|^2+\sigma_{\rm B}^2\left\|\mathbf{z}\right\|^2}$ denotes
the effective SINR received at the FD BS for UL BackCom.
Observing that the optimal solution for receive beamforming is achieved when $\Gamma_{\rm B}$ is maximized, which can be further rewritten as:
\begin{align}
\Gamma_{\rm B}=\frac{\mathbf{z}^H{\boldsymbol{\Upsilon}}{\boldsymbol{\Upsilon}}^H\mathbf{z}}{\mathbf{z}^H\boldsymbol{\lambda}\mathbf{z}},
\end{align}
where $\boldsymbol{\Upsilon}\in\mathbb{C}^{N \times J}$ and $\boldsymbol{\lambda}\in\mathbb{C}^{N \times N}$ are defined as
\begin{gather}
	\boldsymbol{\Upsilon}=\sum_{k\in\mathcal{K}}{\mathbf{H}}_k\widetilde{\mathbf{w}},\\
	\boldsymbol{\lambda}={\mathbf{H}}_{\rm SI}{\mathbf{W}}{\mathbf{H}}_{\rm SI}^H+\sigma_{\rm{B}}^{2} \mathbf{I}_{N},
\end{gather}
respectively. Instead of directly solving problem \eqref{P5}, we resort to the method introduced in~\cite{gershman2010convex} to recast \eqref{P5} as follows:
\begin{subequations}\label{P6}
	\begin{align}	
		&\mathop{\rm{min}}\limits_{\{\mathbf{z}\}}\quad \mathbf{z}^H{\boldsymbol{\lambda}}\mathbf{z}\\
		&\label{P6_C1}\;\;{\rm s.t.} \quad
		\mathbf{z}^H{\boldsymbol{\Upsilon}}=1.
	\end{align}
\end{subequations}
Then, a closed-form solution to problem \eqref{P6} is deduced,
\begin{equation}
	\mathbf{z}^*=\beta\boldsymbol{\lambda}^{-1}\boldsymbol{\Upsilon},
\end{equation}
where $\beta$ is a scalar to adjust $\mathbf{z}^*$ in order to satisfy the equality constraint of \eqref{P6_C1}. We note that for the original subproblem \eqref{P5}, $\beta$ can be omitted since it does not affect the value of the objective function taken.
\subsection{Joint Passive STAR-RIS Beamforming Optimization}
For any given active beamforming vectors $\mathbf{z}$ and $\{\mathbf{w}_j\}$, the subproblem for joint passive beamforming design at the STAR-RIS is reduced into
\begin{subequations}\label{P7}
	\begin{align}	
		&\mathop{\rm{max}}\limits_{\mathbf{ \Theta}_{t},\mathbf{ \Theta}_{r}}\quad  \omega^{\rm UL}{R_{sum}^{\rm UL}}+\omega^{\rm DL}{R_{sum}^{\rm DL}}\\
		&\;\;\;{\rm s.t.} \quad
		\textup{\eqref{P1_C2} -- \eqref{P1_C6}}.
	\end{align}
\end{subequations}
Note that the non-convexity of the obove subproblem steams from the objective function and constraints \eqref{P1_C2}, \eqref{P1_C3} and \eqref{P1_C5}, which makes the reconfiguration for STAR-RIS very difficult.

To facilitate STAR-RIS design, denote the transmission- and reflection-coefficient vectors as $\mathbf{u}_t\triangleq[\sqrt{\beta_1^t}e^{j\theta_{1}^t}, \sqrt{\beta_2^t}e^{j\theta_{2}^t},\cdots, \sqrt{\beta_M^t}e^{j\theta_{M}^t}]^H\in\mathbb{C}^{M \times 1}$ and $\mathbf{u}_r\triangleq[\sqrt{\beta_1^r}e^{j\theta_{1}^r},\sqrt{\beta_2^r}e^{j\theta_{2}^t},\cdots,\sqrt{\beta_M^r}e^{j\theta_{M}^r}]^H\in\mathbb{C}^{M \times 1}$. We first tackle the non-convex objective function in (33a). Specifically, we can rewrite the quadratic term $\left|\mathbf{z}^H\widetilde{\mathbf{h}}_k^H\widetilde{\mathbf{h}}_k\mathbf{x}_0\right|^2$ in the numerator of (7) as $\left|\widetilde{\mathbf{h}}_k\mathbf{z}\right|^2\left|\widetilde{\mathbf{h}}_k\widetilde{\mathbf{w}}\right|^2$. Then, the term $\left|\widetilde{\mathbf{h}}_k\mathbf{z}\right|^2=\left|\mathbf{g}_{k}\mathbf{z}+\mathbf{h}_{{\rm S},k}^H\mathbf \Theta_r\mathbf{f}_{\rm BS}\mathbf{z}\right|^2$ can be explicitly exhibited by the expression in (34), which is shown at the top of the page,
\begin{figure*}[!t]
\normalsize
\begin{equation}
\begin{aligned} \label{eqn2}
	&\left|\mathbf{g}_{k} \mathbf{z}+\mathbf{h}_{{\rm S},k}^H \mathbf \Theta_r \mathbf{f}_{\rm BS} \mathbf{z}\right|^{2}\\ 
	& = \mathbf{g}_{k}\mathbf{Z} \mathbf{g}_{k}^H+2 \Re\left\{\mathbf{g}_{k} \mathbf{Z} \mathbf{f}_{\rm BS}^{H} \mathbf \Theta_r^{H} \mathbf{h}_{{\rm S},k}\right\}+\mathbf{h}_{{\rm S},k}^H \mathbf \Theta_r \mathbf{f}_{\rm BS} \mathbf{Z} \mathbf{f}_{\rm BS}^{H} \mathbf \Theta_r^{H} \mathbf{h}_{{\rm S},k}\\ 
	& = \mathbf{g}_{k} \mathbf{Z} \mathbf{g}_{k}^H+2 \Re\left\{\mathbf{g}_{k} \mathbf{Z} \mathbf{f}_{\rm BS}^{H} \textup{diag}\left(\mathbf{h}_{{\rm S},k}\right) \mathbf{u}_r\right\}+\mathbf{u}_r^{H} \textup{diag}\left(\mathbf{h}_{{\rm S},k}^H\right) \mathbf{f}_{\rm BS} \mathbf{Z} \mathbf{f}_{\rm BS}^{H} \textup{diag}\left(\mathbf{h}_{{\rm S},k}\right) \mathbf{u}_r\\ 
	& = \textup{Tr}\left(\left[\begin{array}{ll}
	\mathbf{u}_r^{H} & \rho^{*}
	\end{array}\right]\left[\begin{array}{c}
		\textup{diag}\left(\mathbf{h}_{{\rm S},k}^H\right) \mathbf{f}_{\rm BS} \\
		\mathbf{g}_{k}
	\end{array}\right]\right. \left. \mathbf{Z}\left[\begin{array}{ll}
		\mathbf{f}_{\rm BS}^{H} \textup{diag}\left(\mathbf{h}_{{\rm S},k}\right) & \mathbf{g}_{k}^H 
	\end{array}\right]\left[\begin{array}{c}
\mathbf{u}_r \\
	\rho
\end{array}\right]\right)\\ 
& = \textup{Tr}\left(\widetilde{\mathbf{u}}_r^{H} \mathbf{Q}_{k} \mathbf{Z} \mathbf{Q}_{k}^{H} \widetilde{\mathbf{u}}_r\right) = \textup{Tr}\left(\mathbf{U}_r \mathbf{Q}_{k} \mathbf{Z} \mathbf{Q}_{k}^{H}\right),
\end{aligned}
\end{equation}
\hrulefill \vspace*{0pt}
\end{figure*}
where $\widetilde{\mathbf{u}}_r \in \mathbb{C}^{(M+1) \times 1}$ and  $\mathbf{U}_r\in \mathbb{C}^{(M+1) \times(M+1)}$ are defined as  $\widetilde{\mathbf{u}}_r\triangleq\left[{\mathbf{u}}_r^{T} \quad \rho\right]^{T}$ and  $\mathbf{U}_r\triangleq\widetilde{\mathbf{u}}_r \widetilde{\mathbf{u}}_r^{H}$, respectively, satisfying $\mathbf U_{r}\succeq 0$ and $\textup{Rank}(\mathbf U_{r})=1$. In addition, $\rho \in \mathbb{C}$  is a dummy variable with  $|\rho|^{2}=1$, $\mathbf{Q}_{k}\triangleq\left[\left(\textup{diag}(\mathbf{h}_{{\rm S},k}^H)\mathbf{f}_{{\rm BS}}\right)^{T} \quad \left({\mathbf{g}_{k}^H}\right)^{*}\right]^{T}\in \mathbb{C}^{(M+1) \times N}$.
Similarly, the term $\left|\widetilde{\mathbf{h}}_k\widetilde{\mathbf{w}}\right|^2=\left|\mathbf{g}_{k}\widetilde{\mathbf{w}}+\mathbf{h}_{{\rm S},k}^H\mathbf \Theta_r\mathbf{f}_{\rm BS}\widetilde{\mathbf{w}}\right|^2$ can be rewritten as $\sum_{j\in\mathcal{J}}\textup{Tr}\left(\mathbf{U}_r \mathbf{Q}_{k} \mathbf{W}_j \mathbf{Q}_{k}^{H}\right)$. Therefore, we can rewrite the UL sum rate function in (7) equivalently as follows:
\begin{equation}
	\begin{aligned} 
	&R_{sum}^{\rm UL}=\\
	&\log_2\left(1+\frac{\sum\limits_{k\in\mathcal{K}}\sum\limits_{j\in\mathcal{J}}\left(\textup{Tr}\left(\mathbf{U}_r \mathbf{Q}_{k} \mathbf{Z} \mathbf{Q}_{k}^{H}\right)\textup{Tr}\left(\mathbf{U}_r \mathbf{Q}_{k} \mathbf{W}_j \mathbf{Q}_{k}^{H}\right) \right)}{\sum\limits_{j\in\mathcal{J}}\rho\textup{Tr}\left(\mathbf W_j\mathbf{H}_{\rm SI}^H\mathbf{Z}\mathbf{H}_{\rm SI}\right)+\sigma_{\rm B}^2\textup{Tr}\left(\mathbf Z\right)}\right).
\end{aligned}
\end{equation}

Then, SCA technique can be applied to construct the concave lower bound of $\textup{Tr}\left(\mathbf{U}_r \mathbf{Q}_{k} \mathbf{Z} \mathbf{Q}_{k}^{H}\right)\textup{Tr}\left(\mathbf{U}_r \mathbf{Q}_{k} \mathbf{W}_j \mathbf{Q}_{k}^{H}\right)$.
To elaborate it, denote $a=\textup{Tr}\left(\mathbf{U}_r \mathbf{Q}_{k} \mathbf{Z} \mathbf{Q}_{k}^{H}\right)$ and $b=\textup{Tr}\left(\mathbf{U}_r \mathbf{Q}_{k} \mathbf{W}_j \mathbf{Q}_{k}^{H}\right)$, with given points $\{\widetilde{a}, \widetilde{b}\}$, we have 
\begin{equation}
	\begin{aligned} 
ab&=\frac{\left(a+b\right)^2-\left(a^2+b^2\right)}{2}\ge[\varphi_{k,j}]^{lb}\\
&\triangleq\left(a+b\right)\left(\widetilde a+\widetilde b\right)-\frac{\left(\widetilde a+\widetilde b\right)^2}{2}-\frac{a^2+b^2}{2}.
\end{aligned}
\end{equation}
By introducing an auxiliary variable $\mu$ satifies
\begin{equation}
	\sum\limits_{k\in\mathcal{K}}\sum\limits_{j\in\mathcal{J}}[\varphi_{k,j}]^{lb}\ge \mu,
\end{equation}
we rewrite the UL sum rate expression in the following concave functional form as
\begin{equation}
	R_{sum}^{\rm UL}=\log_2\left(1+\frac{\mu}{\sum\limits_{j\in\mathcal{J}}\rho\textup{Tr}\left(\mathbf W_j\mathbf{H}_{\rm SI}^H\mathbf{Z}\mathbf{H}_{\rm SI}\right)+\sigma_{\rm B}^2\textup{Tr}\left(\mathbf Z\right)}\right).
\end{equation}

In the following, we aim to obtain a more tractable form of the DL sum rate function. Define $\mathbf{Q}_{k,j}\triangleq\textup{diag}\left(\mathbf{h}_{{\rm S},j}^H\right)\mathbf{h}_{k,{\rm S}}\in\mathbb{C}^{M \times 1}$ and $\mathbf{Q}_j\triangleq\textup{diag}\left(\mathbf{h}_{{\rm S},j}^H\right)\mathbf{f}_{{\rm BS}}\in\mathbb{C}^{M \times N}$, we have $\mathbf{h}_{{\rm S},j}^H\mathbf \Theta_t\mathbf{h}_{k,{\rm S}}=\mathbf{u}_{t}^H\mathbf{Q}_{k,j}$ and
$\mathbf{h}_{{\rm S},j}^H\mathbf \Theta_t\mathbf{f}_{\rm BS}=\mathbf{u}_{t}^H\mathbf{Q}_j$, $\forall k\in\mathcal{K}$, $\forall j\in\mathcal{J}$. The term $\mathcal{E}\left\{I_jI_j^*\right\}$ can be rewritten as
\begin{equation} 
	\begin{aligned}
		\mathcal{E}\left\{I_jI_j^*\right\}&=\sum\limits_{k\in\mathcal{K}}\left(\left|\widetilde{\mathbf{h}}_{k,j}\right|^2\left|\widetilde{\mathbf{h}}_k\widetilde{\mathbf{w}}\right|^2\right)\\
		&=\sum\limits_{k\in\mathcal{K}}{\left(\textup{Tr}\left(\widetilde{\mathbf Q}_{k,j}\mathbf U_{t}\right)\textup{Tr}\left(\mathbf{U}_r \mathbf{Q}_{k} \mathbf{W} \mathbf{Q}_{k}^{H}\right)\right)},
	\end{aligned}
\end{equation}
where $\mathbf{W}\triangleq\sum_{j\in\mathcal{J}}\mathbf{W}_j$, $\widetilde{\mathbf Q}_{k,j}\triangleq\mathbf Q_{k,j}\mathbf Q_{k,j}^H\in \mathbb{C}^{M\times M}$ and $\mathbf{U}_{t}\triangleq\mathbf{u}_{t}\mathbf{u}_{t}^H\in \mathbb{C}^{M \times M}$, respectively, satisfying $\mathbf U_{t}\succeq 0$ and $\textup{Rank}(\mathbf U_{t})=1$.

Let us define ${\mathbf{P}}_{l,j}\triangleq{\mathbf{Q}}_{l}{\mathbf{w}}_j\in\mathbb{C}^{M \times 1}$, $\forall j,l\in \mathcal{J}$. As previously stated, the following two slack variables are introduced to obtain the lower bound of $R_{l\rightarrow j}$: 
 \begin{align}
\frac{1}{X_{lj}}\le&\left|\mathbf{u}_{t}^H{\mathbf{P}}_{l,j}\right|^2=\textup{Tr}\left(\widetilde{\mathbf{P}}_{l,j}\mathbf U_{t}\right),\\
Y_{lj}\ge&\sum\limits_{k\in\mathcal{K}}\left(\textup{Tr}\left(\widetilde{\mathbf Q}_{k,l}\mathbf U_{t}\right)\textup{Tr}\left(\mathbf{U}_r \mathbf{Q}_{k} \mathbf{W} \mathbf{Q}_{k}^{H}\right)\right)\notag\\
&+\sum_{\Omega_i>\Omega_j}\textup{Tr}\left(\widetilde{\mathbf{P}}_{l,i}\mathbf U_{t}\right)+\sigma^2,
\end{align}
where $\widetilde{\mathbf P}_{l,j}\triangleq{\mathbf P}_{l,j}{\mathbf P}_{l,j}^H\in\mathbb{C}^{M \times M}$. 
By applying a relaxation process similar to (21), an approximation of the DL user data rate can be derived as
\begin{equation}
	\begin{aligned}
		\log_2&\left(1+\frac{1}{X_{lj} Y_{lj}}\right)\ge
		\left[R_{l\rightarrow j}\right]^{lb}\\
		\triangleq&\log_2\left(1+\frac{1}{\widetilde X_{lj} \widetilde Y_{lj}}\right)-\frac{X_{lj}-\widetilde X_{lj}}{\ln2\left(\widetilde X_{lj}^2\widetilde Y_{lj}+\widetilde X_{lj}\right)}\\
		&-\frac{Y_{lj}-\widetilde Y_{lj}}{\ln2\left(\widetilde Y_{lj}^2\widetilde X_{lj}+\widetilde Y_{lj}\right)}.
	\end{aligned}
\end{equation}
By denoting $x=\textup{Tr}\left(\widetilde{\mathbf Q}_{k,l}\mathbf U_{t}\right)$ and $y=\textup{Tr}\left(\mathbf{U}_r \mathbf{Q}_{k} \mathbf{W} \mathbf{Q}_{k}^{H}\right)$, with given points $\left\{\widetilde{x}, \widetilde{y}\right\}$, we have
\begin{equation}
	\begin{aligned}
	xy &\le\left[\pi_{k,l}\right]^{ub}\\
	&\triangleq\frac{\left(x+y\right)^2}{2}-\frac{\widetilde{x}^2+\widetilde{y}^2}{2}-{\left(\widetilde{x}\left(x-\widetilde{x}\right)+\widetilde{y}\left(y-\widetilde{y}\right)\right)},
	\end{aligned}
\end{equation}
where $\left[\pi_{k,l}\right]^{ub}$ denotes the convex upper bound of $\textup{Tr}\left(\widetilde{\mathbf Q}_{k,l}\mathbf U_{t}\right)\textup{Tr}\left(\mathbf{U}_r \mathbf{Q}_{k} \mathbf{W} \mathbf{Q}_{k}^{H}\right)$.
Up to this point, by replacing $Y_{lj}$ with $\widehat {Y}_{lj}\triangleq\sum_{k\in\mathcal{K}}\left[\pi_{k,l}\right]^{ub}+\sum_{\Omega_i>\Omega_j}\textup{Tr}\left(\widetilde{\mathbf{P}}_{l,i}\mathbf U_{t}\right)+\sigma^2$ in (42), $\left[R_{l\rightarrow j}\right]^{lb}$ can be further relaxed to a concave functional form, denoted by $[\widehat R_{l\rightarrow j}]^{lb}$. 

Then, problem \eqref{P7} is approximated as the following problem
\begin{subequations}\label{P8}
	\begin{align}
		&\mathop{\rm{max}}\limits_{\{\mathbf{U}_{\chi},{R_{j}^{\rm DL}},X_{lj},Y_{lj}\},\mu}\; \omega^{\rm UL}R_{sum}^{\rm UL}+ \omega^{\rm DL}\sum\limits_{j\in\mathcal{J}}{R_{j}^{\rm DL}}\\
		&\label{P8_C1}{\rm s.t.} \;
		\textup{Tr}\left(\widetilde{\mathbf{P}}_{i,j}\mathbf U_{t}\right)\ge\textup{Tr}\left(\widetilde{\mathbf{P}}_{i,l}\mathbf U_{t}\right), \;{\rm if}\;\Omega_j<\Omega_l,
		\forall i,j,l\in\mathcal{J},\\	
		&\label{P8_C2}\quad\;\;
		\min\left\{[\widehat R_{j\rightarrow j}]^{lb},[\widehat R_{l\rightarrow j}]^{lb}\right\}\ge R_{j}^{\rm DL},
		{\rm if}\;\Omega_j<\Omega_l,\\
		&\label{P8_C3}\quad\;\;
	    \left[\mathbf{U}_t\right]_{mm}+\left[\mathbf{U}_r\right]_{mm}=1,\forall m\in\mathcal{M},\\	
	    &\label{P8_C4}\quad\;\;
	   \mathbf U_{\chi} \succeq 0,\chi\in\{t,r\},\\	
		&\label{P8_C5}\quad\;\;
		\text{Rank}\left(\mathbf{U}_{\chi}\right)=1, \chi\in\{t,r\},\\
		&\quad\;\;
		\textup{\eqref{P1_C4}, \eqref{P1_C6}, (37), (40), (41)}.
	\end{align}
\end{subequations}
We note that the non-convex rank-one constraint also appears in problem \eqref{P8}, which is challenging to deal with.
After dropping constraint \eqref{P8_C5}, the resulting problem can be solved directly by using the the semidefinite relaxation (SDR) technique~\cite{luo2010semidefinite}, and the solution obtained by this relaxation problem serves as an upper bound of the optimal solution for problem \eqref{P8}.

Here is an expression equivalent to the rank-one constraint, such as
\begin{equation}
	\left\|\mathbf{U}_{\chi}\right\|_*-\left\|\mathbf{U}_{\chi}\right\|_2=0, \chi\in\left\{t,r\right\},
\end{equation}
where $\left\|\mathbf{U}_{\chi}\right\|_*\triangleq\sum_i\sigma_i\left(\mathbf{U}_{\chi}\right)$ and $\left\|\mathbf{U}_{\chi}\right\|_2\triangleq \sigma_1\left(\mathbf{U}_{\chi}\right)$ denote the nuclear norm and spectral norm, respectively, and $\sigma_i\left(\mathbf{U}_{\chi}\right)$ is the $i$-th largest singular value of matrix $\mathbf{U}_{\chi}$. By incorporating the equality constraint (45) into the objective function of problem \eqref{P8}, the problem can be reformulated as
\begin{subequations}\label{P9}
	\begin{align}
		&\min _{\left\{\mathbf{U}_{\chi}, R_{j}^{\mathrm{DL}}, X_{l j}, Y_{l j}\right\}, \mu}\quad
		\begin{array}{l}
		-\left(\omega^{\rm UL}R_{sum}^{\rm UL}+\omega^{\rm DL}\sum_{j\in\mathcal{J}}{R_{j}^{\rm DL}}\right) \\
		+\frac{1}{\eta}  \sum_{\chi\in\{t,r\}}\left(\left\|\mathbf{U}_{\chi}\right\|_*-	\left\|\mathbf{U}_{\chi}\right\|_2\right)
		\end{array}\\
		&\quad\;\;{\rm s.t.} \quad
		\textup{\eqref{P1_C4}, \eqref{P1_C6}, (37), (40), (41), \eqref{P8_C1} -- \eqref{P8_C4}},
	\end{align}
\end{subequations}
where $\eta>0$ is the penalty factor which penalizes the objective function if $\{\mathbf{U}_{\chi}\}$ is not rank-one. 
Fortunately, it is revealed that the rank-one property of the optimal solution $\{\mathbf{U}_{\chi}^{\ast}\}$ to problem \eqref{P9} can always be guaranteed when $\eta\to 0$. 

Since the penalty term is a difference of two convex functions~\cite{ben1997penalty}, we resort to the SCA technique to approximately obtain its convex upper bound expression, as shown below
\begin{equation} 
	\begin{split}
		\left\|\mathbf{U}_{\chi}\right\|_*-	\left\|\mathbf{U}_{\chi}\right\|_2 \le\left\|\mathbf{U}_{\chi}\right\|_*-\overline{\mathbf U}_{\chi}, \forall\chi\in\left\{t,r\right\},
	\end{split}
\end{equation} 
where $\overline{\mathbf U}_{\chi}\triangleq\left\|\widetilde{\mathbf U}_{\chi}\right\|_2+\textup{Tr}\left(\partial_{\rm max}\left(\widetilde{\mathbf U}_{\chi}\right)\partial_{\rm max}\left(\widetilde{\mathbf U}_{\chi}\right)^H\left(\mathbf{U}_{\chi}-\widetilde{\mathbf U}_{\chi}\right)\right)$. $\widetilde{\mathbf U}_{\chi}$ denotes the given local point, $\partial_{\rm max}\left(\widetilde{\mathbf U}_{ \chi}\right)$ denotes the eigenvector corresponding to the largest eigenvalue of $\widetilde{\mathbf U}_{\chi}$. 
Till now, the optimization problem can be reformulated as
\begin{subequations}\label{P10}
	\begin{align}
		&\min _{\left\{\mathbf{U}_{\chi}, R_{j}^{\mathrm{DL}}, X_{l j}, Y_{l j}\right\}, \mu}\quad
		\begin{array}{l}
			-\left(\omega^{\rm UL}R_{sum}^{\rm UL}+\omega^{\rm DL}\sum_{j\in\mathcal{J}}{R_{j}^{\rm DL}}\right) \\
			+\frac{1}{\eta}  \sum_{\chi\in\{t,r\}}\left(\left\|\mathbf{U}_{\chi}\right\|_*-	\overline{\mathbf U}_{\chi}\right)
		\end{array}\\
		&\quad\;\;{\rm s.t.} \quad
		\textup{\eqref{P1_C4}, \eqref{P1_C6}, (37), (40), (41), \eqref{P8_C1} -- \eqref{P8_C4}}.
	\end{align}
\end{subequations}
The above problem is convex and therefore standard optimisation tools can be applied to obtain its global optimal solution. In addition, the details for solving problem \eqref{P10} with the penalty based rank-one relaxation method are summarized in \textbf{Algorithm 1}. 
\begin{algorithm}[t]\label{Algorithm 1}
	\caption{The Proposed Penalty Based Algorithm for Solving Problem \eqref{P10}.}
	\begin{algorithmic}[1]
		\STATE \textbf{Initialize} $\{{\mathbf{U}_{\chi}^{(0)}}\}$, $\forall\chi\in\{t,r\}$, $\eta$, $r_1=0$, $r_2=0$,  $\epsilon$, $\varepsilon$, $r_{max}$.
    	\STATE {\bf repeat: outer loop}
		\STATE \quad {\bf repeat: inner loop}
		\STATE \quad\quad Update $\{{\mathbf{U}_{\chi}^{(r_1+1)}}\}$ by solving problem \eqref{P10} with given $\{\mathbf{W}_j\}$ and $\mathbf{z}$. 
		\STATE \quad\quad $r_1\leftarrow r_1+1$.
		\STATE \quad {\bf until} the fractional decrease of the objective function value is below a predefined threshold $ \varepsilon >0$ or the maximum number of inner iterations $r_{\max}$ is reached.
		\STATE \quad Update $\eta^{(r_2)}\leftarrow c\eta^{(r_2)}$.
		\STATE \quad $r_2\leftarrow r_2+1$ and $r_1\leftarrow0$.
		\STATE {\bf until} $\max\left\{ \left\|\mathbf{U}_{\chi}\right\|_*-\left\|\mathbf{U}_{\chi}\right\|_2, \forall\chi\in\{t,r\}\right\}\le \epsilon$, where $\epsilon>0$ is a predefined threshold for constraint violation.
	\end{algorithmic}
\end{algorithm}
\subsection{Proposed Algorithm and Complexity Analysis}
For a given decoding order, by solving subproblems \eqref{P4}, \eqref{P6} and subproblem \eqref{P10}, the FD BS active beamforming coefficients $\{\mathbf{w}_j\}$ and $\mathbf{z}$ and STAR-RIS passive beamforming coefficients $\left\{\mathbf{\Theta}_{\chi},\forall\chi\in\left\{t,r\right\}\right\}$ are optimized alternately. We note that the receive beamforming vector $\mathbf{z}^*$ at FD BS admits a closed-form solution, cf. (32). Moreover, the overall AO based algorithm is summarized in \textbf{Algorithm 2}. 
As the objective function is a non-decreasing function in each iteration and is limited by an upper bound, it is guaranteed to eventually converge~\cite{dinh2010local}.

In the following, the computational complexity of \textbf{Algorithm 2} for soving problem \eqref{P1} is analyzed. Firstly, the complexity to design the FD BS transmit beamforming in problrm \eqref{P4} is in order of $\mathcal{O}\left(\max \left(3J^{2}+2J+4, N\right)^{4} \sqrt{N} \log _{2}(1 / \varepsilon)\right)$~\cite{luo2010semidefinite}. We note that the main complexity of \textbf{Algorithm 1} steams from solving the relaxed problem \eqref{P10} in the inner layer iteration. Specifically, the complexity to deal with this standard SDP for STAR-RIS transmission- and reflection-coefficient optimization is in order of  $\mathcal{O}\left(\max \left(4J^{2}+J+1, M\right)^{4} \sqrt{M} \log _{2}(1 / \varepsilon)\right)$.
In consequence, the total complexity of \textbf{Algorithm 2} is ${{\mathcal{O}}}(I_{ite}(\max \left(3J^{2}+2J+4, N\right)^{4} \sqrt{N}+I_{ out}I_{inn}(\max \left(4J^{2}+J+1, M\right)^{4} \sqrt{M}))\log _{2}(1 / \varepsilon))$, where $I_{inn}$ and $I_{out}$ respectively denote the number of iterations for reaching convergence in the inner layer and outer layer required by  \textbf{Algorithm 1}, respectively, $I_{ite}$ denotes the number of iterations required by \textbf{Algorithm 2}.
\begin{algorithm}[t]\label{Algorithm 2}
	\caption{The Proposed Alternating Optimization Based Algorithm for Solving Problem \eqref{P1}.}
	\label{alg:A}
	\begin{algorithmic}[1]
		\STATE \textbf{Initialize} $\{\mathbf{W}_j^{(0)}\}$, $\mathbf{z}^{(0)}$ and $\{\mathbf{U}_{\chi }^{(0)}\},\forall\chi\in\{t,r\}$, $n=0$, $ \varepsilon$.
		\REPEAT 
		\STATE {
			Update $\{\mathbf{W}_j^{(n+1)}\}$} by solving problem \eqref{P4}.
		\STATE {
			Calculate the optimal FD BS receive beamforming coefficient by using the closed-form expression in (32), and update ${\mathbf{z}^{(n+1)}}$} .
		\STATE {Solve problem \eqref{P10} with the proposed \textbf{Algorithm 1}, and update $\{\mathbf{U}_{\chi }^{(n+1)}\}$}.\\
		\STATE $n \leftarrow n+1$.\\
		\UNTIL the fractional increase of the objective value is below a threshold $\varepsilon$.\\
		\STATE	\textbf{Output} the optimal solutions ${\mathbf{W}_j^{(*)}}={\mathbf{W}_j^{(n)}}$, ${\mathbf{z}^{(*)}}={\mathbf{z}^{(n)}}$ and ${\mathbf{U}_{\chi }^{(*)}}={\mathbf{U}_{\chi }^{(n)}}$.
	\end{algorithmic}
\end{algorithm}
\section{Low-Complexity DL NOMA Decoding Order Scheme}
In NOMA systems, it is particularly important to develop a proper decoding order based on the strength of the user channels to eliminate inter-user interference induced by resource multiplexing~\cite{liu2022evolution}. Unlike UL NOMA transmissions, the DL system sum rate is directly determined by the SIC decoding order. 
Therefore, the decoding order of the DL users needs to be determined before solving the original optimization problem. Generally, for the $J$ DL users, we can solve the original problem $J!$ times to select the optimal decoding order. Nevertheless, the complexity of the exhaustive search method is prohibitive, e.g., ${{\mathcal{O}}}(J!)$, especially when $J$ is large. To fully reap the potential of DL NOMA transmission with affordable complexity, we propose a low-complexity scheme in the following to optimize the user decoding orders by maximizing the sum of the combined channel gains. As a result, the decoding order optimization problem is formulated as follows:
\begin{subequations}\label{P11}
	\begin{align}	
		&\mathop{\rm{max}}\limits_{\{\mathbf{w}_j\},\mathbf{ \Theta}_t}\quad\sum_{j\in\mathcal{J}}\left|(\mathbf{h}_{{\rm S},j}^H\mathbf \Theta_t\mathbf{f}_{\rm BS})\mathbf{w}_j\right|^2\\
		&\label{P11_C1}\;\quad{\rm s.t.} \quad
		\sum_{j\in\mathcal{J}}\left\|\mathbf{w}_j\right\|^2\le P_{\max},\\
		&\label{P11_C2}\quad\quad\quad\;\;
		\left[\mathbf{\Theta}_t\right]_{m,m} \leq 1, \forall m \in \mathcal{M}.
	\end{align}
\end{subequations}
Before solving problem \eqref{P11}, we claim that the combined channel gains of DL user $j\in\mathcal{J}$ can be given by
\begin{equation}
	\phi_j=\left|(\mathbf{h}_{{\rm S},j}^H\mathbf \Theta_t\mathbf{f}_{\rm BS})\mathbf{w}_j\right|^2=\textup{Tr}\left(\mathbf{U}_t\mathbf{Q}_j\mathbf{W}_j\mathbf{Q}_j^H\right).
\end{equation}
Therefore, problem \eqref{P11} can be transformed into a more tractable form as
\begin{subequations}\label{P12}
	\begin{align}	
		&\mathop{\rm{max}}\limits_{\{\mathbf{W}_j\},\mathbf{U}_t}\quad\sum_{j\in\mathcal{J}}\textup{Tr}\left(\mathbf{U}_t\mathbf{Q}_j\mathbf{W}_j\mathbf{Q}_j^H\right)\\
		&\label{P12_C1}\;\quad{\rm s.t.}\quad
		\sum_{j\in\mathcal{J}}\textup{Tr}\left(\mathbf{W}_j\right)\le P_{\max},\\
		&\label{P12_C2}\quad\quad\quad\;\;\,
		\left[\mathbf{U}_t\right]_{m,m} \leq 1, \forall m \in \mathcal{M},\\
		&\label{P12_C3}\quad\quad\quad\;\;\,
		\mathbf{W}_j \succeq 0, \mathbf{U}_t \succeq 0,\\
		&\label{P12_C4}\quad\quad\quad\;\;\,
		\textup{Rank}\left(\mathbf{W}_j\right)=1,
		\textup{Rank}\left(\mathbf{U}_t\right)=1.
	\end{align}
\end{subequations}
Next, we exploit the following lemma to decompose the coupled variables between $\{\mathbf W_j\}$ and ${\mathbf U}_t$ in the objective function of problem \eqref{P12}.
\begin{lemma}\label{AB}
	\emph{For any two Hermitian matrices ${\mathbf{A}} \in {{\mathbb{H}}^M}$ and ${\mathbf{B}} \in {{\mathbb{H}}^M}$ having the same size, we have the following equality:}
	\begin{align}\label{D 1}
		{\textup{Tr}}\left( {{\mathbf{AB}}} \right)= \frac{1}{4}\left(\underbrace{\left\| {{\mathbf{A}} + {\mathbf{B}}} \right\|_F^2}_{a_1} - \underbrace{\left\| {{\mathbf{A}}- {\mathbf{B}}} \right\|_F^2}_{a_2}\right).
	\end{align}
\end{lemma}
Based on \textbf{Lemma 1}, by leveraging the first-order Taylor expansion of $a_1$ and $a_2$, the concave lower bound expressions of ${\textup{Tr}}\left( {{\mathbf{AB}}} \right)$ are respectively given by
\begin{equation}
	\begin{aligned}	
	{\textup{Tr}}&\left( {{\mathbf{AB}}} \right)\ge{\Pi _{\mathbf{A,B}}^{lb}}\\\triangleq&\frac{1}{4}\bigg\{2{\textup{Tr}}\left(\left( {{\mathbf{A+B}}} \right)\left({\widetilde{\mathbf{A}}+\widetilde{\mathbf{B}}}\right)^H\right)-\left\|\widetilde{\mathbf A}+\widetilde{\mathbf B}\right\|_F^2\\
	&-\left\|{\mathbf A}-{\mathbf B}\right\|_F^2\bigg\},
		\end{aligned}	
\end{equation}
where $\widetilde{\mathbf A}$ and $\widetilde{\mathbf B}$ denote the local points generated in the previous iteration. Then, a lower bound of $\textup{Tr}(\mathbf{U}_t\mathbf{Q}_j\mathbf{W}_j\mathbf{Q}_j^H)$ is obtained as
\begin{equation}
	\begin{aligned}	
	{[\phi_j]}^{lb}\triangleq&\frac{1}{4}\bigg\{2{\textup{Tr}}\left(\left({{{\mathbf U_t+{\mathbf{Q}}_j\mathbf{W}_j{\mathbf{Q}}_j^H}}} \right)\left({{{\widetilde{\mathbf U}_t+{{\mathbf{Q}}_j\widetilde{\mathbf{W}}_j{\mathbf{Q}}_j^H}}}}\right)^H\right)\\
	&-\left\|{{{\widetilde{\mathbf U}_t+{{\mathbf{Q}}_j\widetilde{\mathbf{W}}_j{\mathbf{Q}}_j^H}}}}\right\|_F^2-\left\|{{{\mathbf U_t-{\mathbf{Q}}_j\mathbf{W}_j{\mathbf{Q}}_j^H}}} \right\|_F^2\bigg\}.
		\end{aligned}	
\end{equation}
For given feasible points $\left(\widetilde{\mathbf W}_j, \widetilde{\mathbf U}_t\right)$, we can rewrite the optimization problem for maximizing the sum of the DL system combined channel gains as follows:
\begin{subequations}\label{P13}
	\begin{align}	
		&\mathop{\rm{max}}\limits_{\{\mathbf{W}_j\},\mathbf{U}_t}\quad\sum_{j\in\mathcal{J}}[\phi_j]^{lb}\\
		&\;\quad{\rm s.t.} \quad
		\textup{\eqref{P12_C1} -- \eqref{P12_C4}}.
	\end{align}
\end{subequations}
For the solution of this problem, the only challenge is caused by the rank-one constraint \eqref{P12_C4}. 
To this end, we adopt SDR and solve the relaxed problem without \eqref{P12_C4}. The tightness of ignoring $\textup{Rank}(\mathbf{W}_j )=1$ for joint active and passive beamforming design is given in the following theorem.
\begin{theorem}\label{rank one relax}
	\emph{Without loss of optimality, for the relaxed version of the problem in \eqref{P13}, i.e. without the rank-one constraint \eqref{P12_C4}, the optimal solution $\mathbf{W}_j^*$ obtained always satisfies $\textup{Rank}(\mathbf{W}_j )=1, \forall j\in\mathcal{J}$.}
	\begin{proof}
		\emph{See Appendix A.}
	\end{proof}
\end{theorem}

Armed by the above discussion, the Gaussian randomization technique~\cite{luo2010semidefinite} can be further adopted to obtain a suboptimal solution for satisfying $\textup{Rank}(\mathbf{U}_t)=1$. Next, we order $J$ DL users in descending order of the combined channel gains. In particular, the proposed user order criterion can be specified as 
\begin{equation}
	\begin{aligned}
		\Omega_j<\Omega_l,\quad&{\rm if}\; \textup{Tr}\left(\mathbf{U}_t\mathbf{Q}_j\mathbf{W}_j\mathbf{Q}_j^H\right)\le\textup{Tr}\left(\mathbf{U}_t\mathbf{Q}_l\mathbf{W}_l\mathbf{Q}_l^H\right),\\
		&\forall j,l\in\mathcal{J}.
	\end{aligned}
\end{equation}
The proposed algorithm for decoding order optimization is summarized in \textbf{Algorithm 3}.

\begin{algorithm}[t]\label{Algorithm 3}
	\caption{The Proposed Algorithm for Decoding Order Optimization.}
	\label{alg:A}
	\begin{algorithmic}[1]
		\STATE \textbf{Initialize} $\{\mathbf{W}_j^{(0)}\}$ and $\{\mathbf{U}_{t }^{(0)}\}$, $n=0$, $ \varepsilon$.
		\REPEAT 
		\STATE 
			Update $\{\mathbf{w}_j^{(n+1)}\}$ and $\{\mathbf{U}_{t }^{(n+1)}\}$ by solving problem \eqref{P13}.
		\STATE $n \leftarrow n+1$.\\
		\UNTIL the fractional increase of the objective value is below a threshold $\varepsilon$.\\	
		\STATE Calculate $\left\{\textup{Tr}\left(\mathbf{U}_t\mathbf{Q}_j\mathbf{W}_j\mathbf{Q}_j^H\right), \forall  j\in\mathcal{J}\right\}$ and then order $J$ DL users in descending order of the combined channel gains.\\
		\STATE	\textbf{Output} the decoding orders $\Omega_j,  \forall  j\in\mathcal{J}$.
	\end{algorithmic}
\end{algorithm}
\section{Numerical Results}
In this section, simulation results are presented to evaluate the effectiveness of the proposed STAR-RIS assisted DL active and UL backscatter communications based on NOMA protocol. 
\subsection{Simulation Setup}
\begin{figure}[t]
	\centering
	\setlength{\belowcaptionskip}{+0.2cm}  
	\includegraphics[width=3.4in]{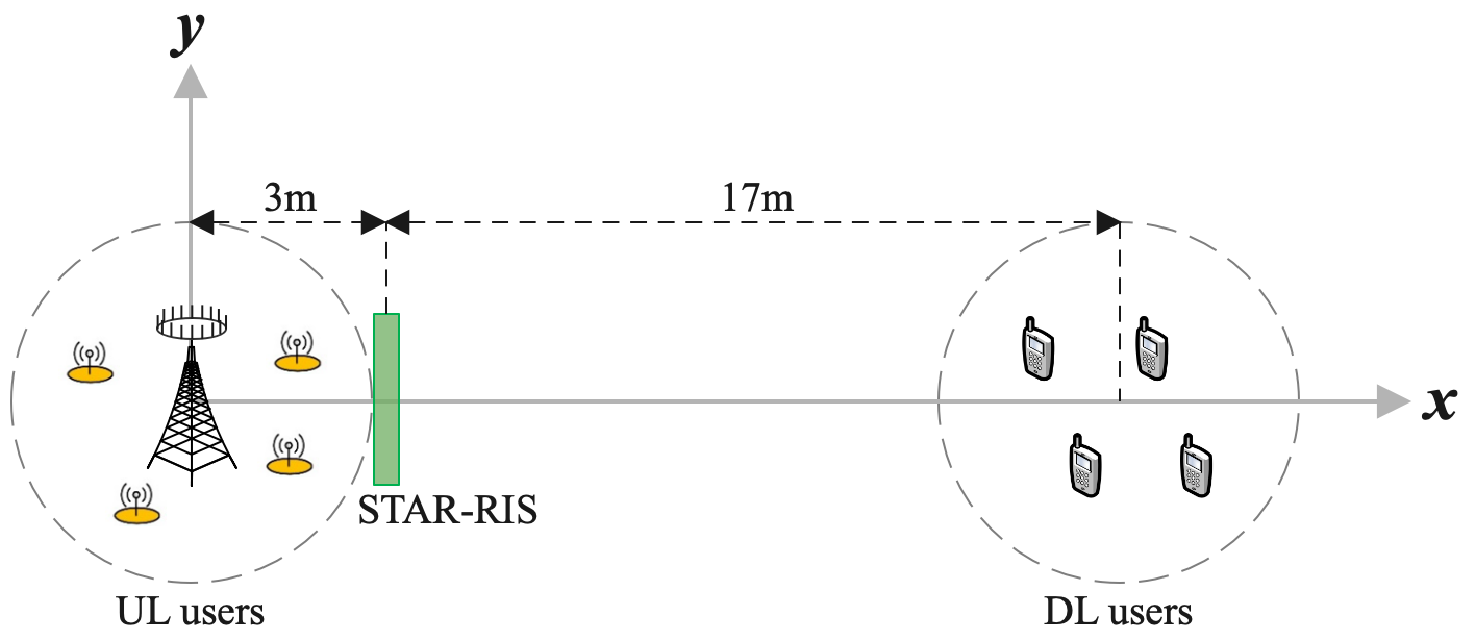}\\
	\caption{The simulated setup.}\label{setup}
\end{figure}
\begin{table*}[t]\small
	\centering
	\caption{System Parameters}
	\begin{tabular}{|l|l|}
		\hline
		\centering
		Path loss at the reference distance of 1 meter  & $\rho_0=-30$ dB \\
		\hline
		\centering
		Rician factor for the STAR-RIS-assisted channels  & $\kappa_{\rm BS}=\kappa_{\rm SU}=3$ dB \\ 
		\hline
		\centering
		Path-loss exponents for the direct channels and the STAR-RIS-assisted channels   & $\alpha_{\rm BU}=3.5$, $\alpha_{\rm BS}= \alpha_{\rm SU}= 2.2$    \\ 
		\hline
		\centering
		Noise power at the receiver  & $\sigma^2 =\sigma_{\rm B}^2= -110$ dBm  \\ 
		\hline
		\centering
		SI cancellation coefficient at the FD BS  & $\rho= -100$ dB~\cite{elhattab2021reconfigurable} \\ 
		\hline
		\centering
		Initialized penalty factor for \textbf{Algorithm 1}   & $\eta={10^{ - 4}}$   \\ 
		\hline
		\centering
		Maximum number of inner iterations for \textbf{Algorithm 1}   & $r_{\max}=30$   \\ 
		\hline		
		\centering
		Convergence accuracy   & ${{\varepsilon}}={10^{ -3}}, {\epsilon}={10^{ - 7}}$  \\ 
		\hline
	\end{tabular}
	\centering
	\vspace{-0.5cm}
\end{table*}
 In Fig. 2, the schematic simulation setup for the considered system model is illustrated, where the BS and the STAR-RIS are located at $\textup{(0 m, 0 m, 0 m)}$ and $\textup{(3 m, 0 m, 0 m)}$, respectively. 
The UL BackCom users and DL users are uniformly and randomly distributed in two circular regions with radius $r=2$ m centered on the FD BS and $\textup{(20 m, 0 m, 0 m)}$, respectively.
Then, the direct links between FD BS and UL BackCom users are modeled as Rayleigh fading channels, and all the links assisted by the STAR-RIS follow Rician fading, which are given by:
\begin{align}
	&\mathbf{g}=\sqrt{PL(d_{\rm BU})}\mathbf{g}^{\rm NLoS},\\
	&\mathbf{f}=\sqrt{PL(d_{\rm BS})}\Bigg(\sqrt{\frac{\kappa_{\rm BS}}{\kappa_{\rm BS}+1}}\mathbf{f}^{\rm LoS}+\sqrt{\frac{1}{\kappa_{\rm BS}+1}}\mathbf{f}^{\rm NLoS}\Bigg),\\	&\mathbf{h}=\sqrt{PL(d_{\rm SU})}\Bigg(\sqrt{\frac{\kappa_{\rm SU}}{\kappa_{\rm SU}+1}}\mathbf{h}^{\rm LoS}+\sqrt{\frac{1}{\kappa_{\rm SU}+1}}\mathbf{h}^{\rm NLoS}\Bigg),
\end{align}
where ${PL(d_{\rm BU})}$, ${PL(d_{\rm BS})}$ and ${PL(d_{\rm SU})}$ are the distance-dependent path loss following $PL(d)=\rho_0 (d)^{-\alpha}$, in which $\rho_0$ denotes the path loss at the reference distance $1$ m, $d$ and $\alpha$  denote the distance and the path loss exponent of the communication link. 
$\kappa_{\rm BS}$ and $\kappa_{\rm SU}$ denote the Rician factors, $\mathbf{f}^{\rm LoS}$ and $\mathbf{h}^{\rm LoS}$ are the deterministic line-of-sight (LoS) components, $\mathbf{g}^{\rm NLoS}$, $\mathbf{f}^{\rm NLoS}$ and $\mathbf{h}^{\rm NLoS}$ are the random nonline-of-sight (NLoS) components modeled as Rayleigh fading. The primary simulation parameters are set out in Table I for convenience of the readers~\cite{mu2021simultaneously,elhattab2021reconfigurable}.
Without loss of generality, in the figures that follow the maximum power allowance for the FD BS is set to be $P_{\max} = 5$ W.  All DL users have identical target rate requirement, i.e., $\overline R=0.1$ bit/s/Hz. The antenna number of the FD BS is $N = 3$ and the STAR-IRS is equipped with a uniform planar array (UPA) consisting of $M=20$ elements. Moreover, we assume a symmetrical number of UL and DL users, i.e., $I_{user}=K=J=4$. For a fair comparison, the system performance of the proposed transmission design is studied with $\omega^{\rm UL}=\omega^{\rm DL}=1$.
\subsection{Convergence of the Proposed Algorithms}
\begin{figure}[t]
	\centering
	\setlength{\belowcaptionskip}{+0.2cm}  
	\includegraphics[width=3.3in]{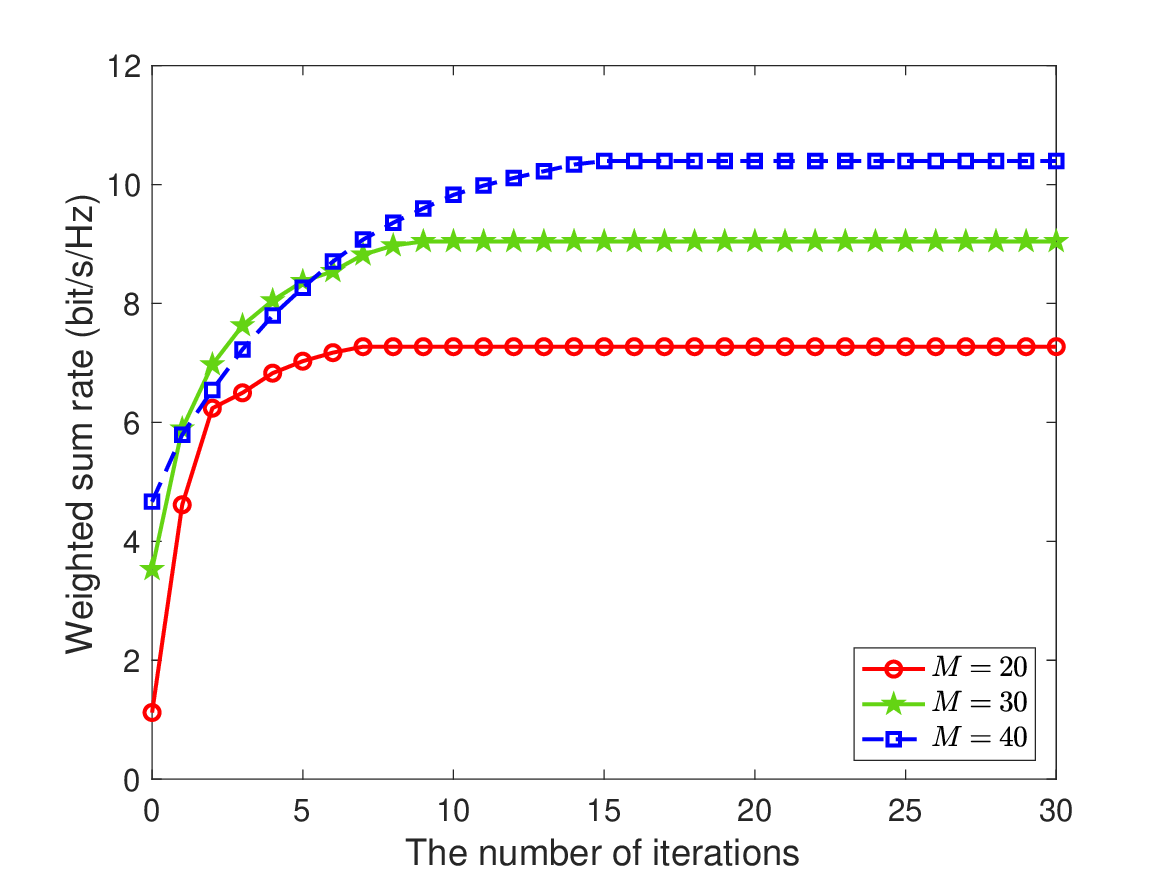}
	\caption{Convergence of \textbf{Algorithm 2}.}
	\label{algorithm convergence}
\end{figure}
Fig. 3 shows the convergence behavior of the proposed  \textbf{Algorithm 2}, by plotting the weighted sum rate versus the number of iterations. 
The proposed algorithm is guaranteed to converge whithin a few iterations for schemes considering different number of STAR-RIS elements.
It is worth noting that, the obtained performance gain increases with the growing $M$, which indicates that more elements on the STAR-RIS contribute to a higher combined channel gain. We also note that the number of elements directly affects the number of iterations required for the algorithm to reach convergence. As can be seen from the figure, more iterations are required by the proposed scheme with larger $M$. This is because the larger arrays of STAR-RIS make the solution of the problem more complicated.
\subsection{Baseline Schemes}
To better investigate the performance achieved by the proposed system design, the following three baseline schemes are considered for comparison.
\begin{itemize}
	\item \textbf{Conventional RIS assisted NOMA scheme (also referred to as CR-NOMA scheme)}: 
	 In this case, both DL active and UL passive radio communications employ the NOMA protocol for transmission. While, the STAR-RIS operates in a conventional RIS mode, where $M/2$ elements are used to transmit signals only, and $M/2$ elements are used to reflect signals only to facilitate \textit{full-space} coverage. The transmission- and reflection-coefficients of the conventional RIS is set to be ${\bm{\beta}_t} \triangleq\left[{\beta}_1^{t},{\beta}_2^{t},\cdots,{\beta}_M^{t}\right]=[{{\mathbf{1}}_{1 \times {M \mathord{\left/{\vphantom {M 2}} \right.\kern-\nulldelimiterspace} 2}}}\;{{\mathbf{0}}_{{{1 \times M} \mathord{\left/{\vphantom {{1 \times M} 2}} \right.\kern-\nulldelimiterspace} 2}}}]$ and ${\bm{\beta}_r} \triangleq\left[{\beta}_1^{r},{\beta}_2^{r},\cdots,{\beta}_M^{r}\right]= [{{\mathbf{0}}_{{{1 \times M} \mathord{\left/{\vphantom {{1 \times M} 2}} \right.\kern-\nulldelimiterspace} 2}}}\;{{\mathbf{1}}_{1 \times {M \mathord{\left/{\vphantom {M 2}} \right.\kern-\nulldelimiterspace} 2}}}]$.
	 \item \textbf{STAR-RIS assisted NOMA-SDMA scheme (also referred to as SR-NOMA-SDMA scheme)}: In this case, the STAR-RIS operates in the ES mode, with all elements can simultaneously transmit and reflect signals. We assume that on the one hand, the NOMA protocol is employed to support connectivity for UL passive BackCom. On the other hand, SDMA is exploited in the DL active communication system.
	 \item 	\textbf{STAR-RIS assisted ZF-NOMA scheme (also referred to as SR-ZF-NOMA scheme)}: In this case, the STAR-RIS operates in the ES mode, with all elements can simultaneously transmit and reflect signals. We assume that on the one hand, the NOMA technique is applied for DL active transmission, and each receiver decodes message via SIC. On the other hand, the FD BS capitalizes on ZF detection to decode the backscattered signals from the UL users.
\end{itemize}
\subsection{Impact of the Number of STAR-RIS Elements}
\begin{figure}[t]
	\centering
	\setlength{\belowcaptionskip}{+0.2cm} 
	\includegraphics[width=3.3in]{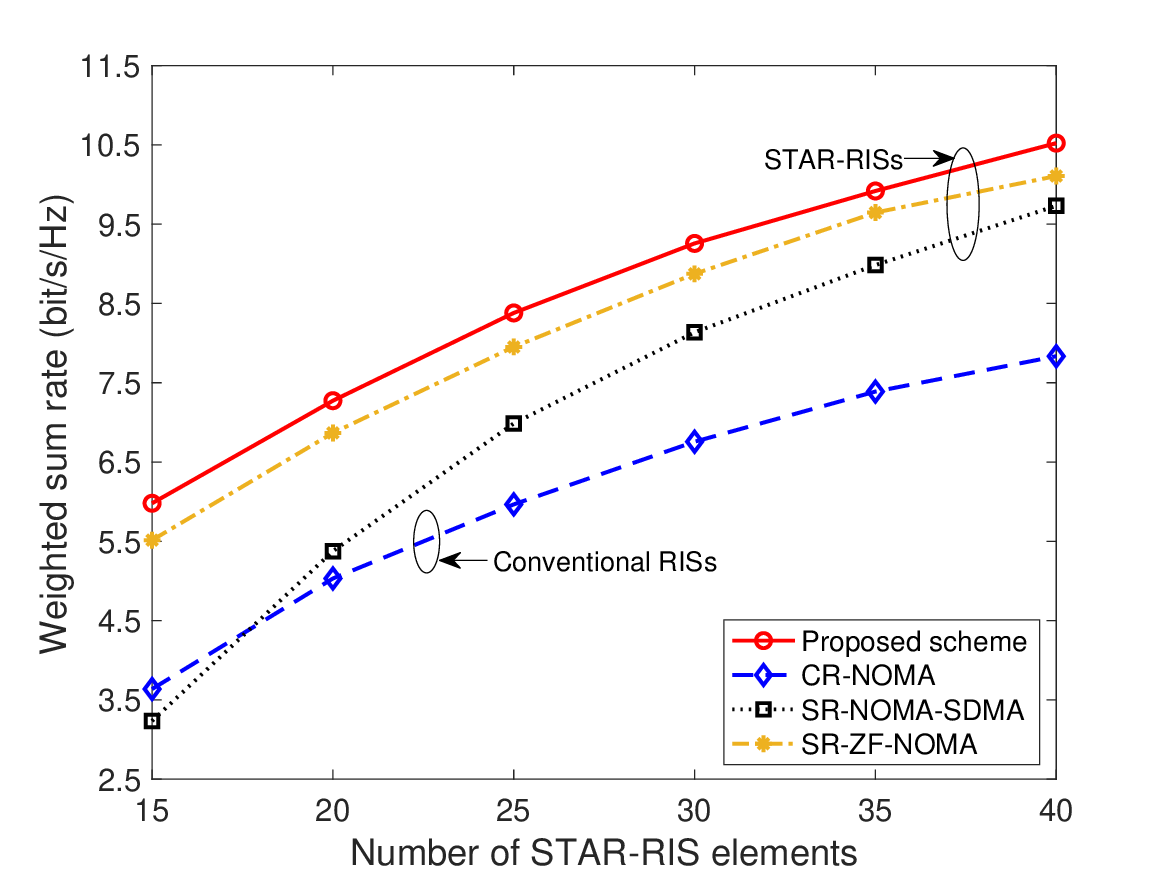}
	\caption{System weighted sum rate versus the number of STAR-RIS element $M$. }
	\label{STAR-RIS elements}
\end{figure}
Fig. 4 presents the system weighted sum rate versus the number of the reflection/transmission elements of the STAR-RIS. As shown in this figure, our proposed scheme always outperforms other baseline schemes. By flexibly adjusting the coefficients of each element on STAR-RIS, the reflection and transmission of the incident signals can be achieved simultaneously. Unlike conventional RIS that can only reflect or transmit signals, STAR-RIS provides additional DoFs to achieve enhanced signal coverage.
In addition, as observed from Fig. 4, SR-ZF-NOMA scheme suffers from some performance degradation as compared to our proposed design. Although ZF beamforming permits good suppression on multiuser interference, it also amplifies the noise and SI at the FD BS, limiting the performance gain of the system. By comparing with SR-NOMA-SDMA scheme, utilizing NOMA protocol in both DL active and UL passive backscatter communications encourages spectrum cooperation among more users. Moreover, the SIC technique is applied to subtract interference from the received signal at the receiver, which is beneficial to enhance the desired signal strength and thus improve the system performance.
\subsection{Impact of the SI cancellation coefficient at the FD BS}
\begin{figure}[t]
	\centering
	\setlength{\belowcaptionskip}{+0.2cm}  
	\includegraphics[width=3.3in]{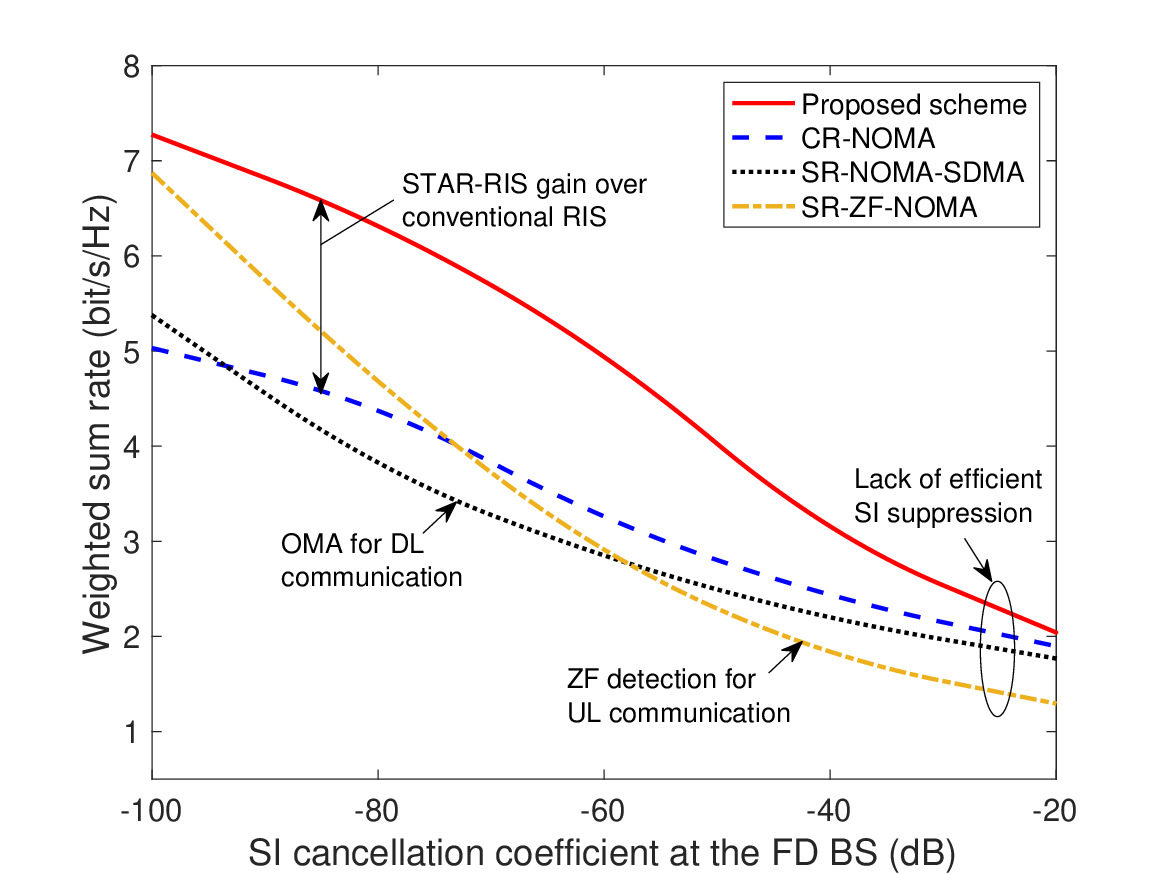}
	\caption{System weighted sum rate versus the SI cancellation coefficient at the FD BS. }
	\label{SI cancellation coefficient}
\end{figure}
In Fig. 5, the system weighted sum rate versus the SI cancellation coefficient at the FD BS is provided.
It can be seen that the performance of all schemes is degraded when SI signal becomes more severe. In fact, the increase of residual SI has the most devastating impact on the UL signal reception in SR-NOMA-SDMA scheme, which is consistent to the conclusion obtained for Fig. 4.
Also, Fig. 5 indicates that the proposed scheme outperforms other schemes, particularly in the case with a low $\rho$, i.e., where the SI signal power is significantly reduced by the SI cancellation technique.
This is because the efficient SI suppression enables the system to well manage the wireless resources, thus making the superiority of our designed transmission scheme even more pronounced. Otherwise, the DoFs for system beamforing design will be diminished and the system performance will also considerably deteriorate in our proposed scheme. Based on these observations, we can conclude that the powerful SI cancellation is important for FD transmissions.
\subsection{DL-UL Sum Rate Tradeoff}
\begin{figure}[t]
	\centering
	\setlength{\belowcaptionskip}{+0.2cm} 
	\includegraphics[width=3.3in]{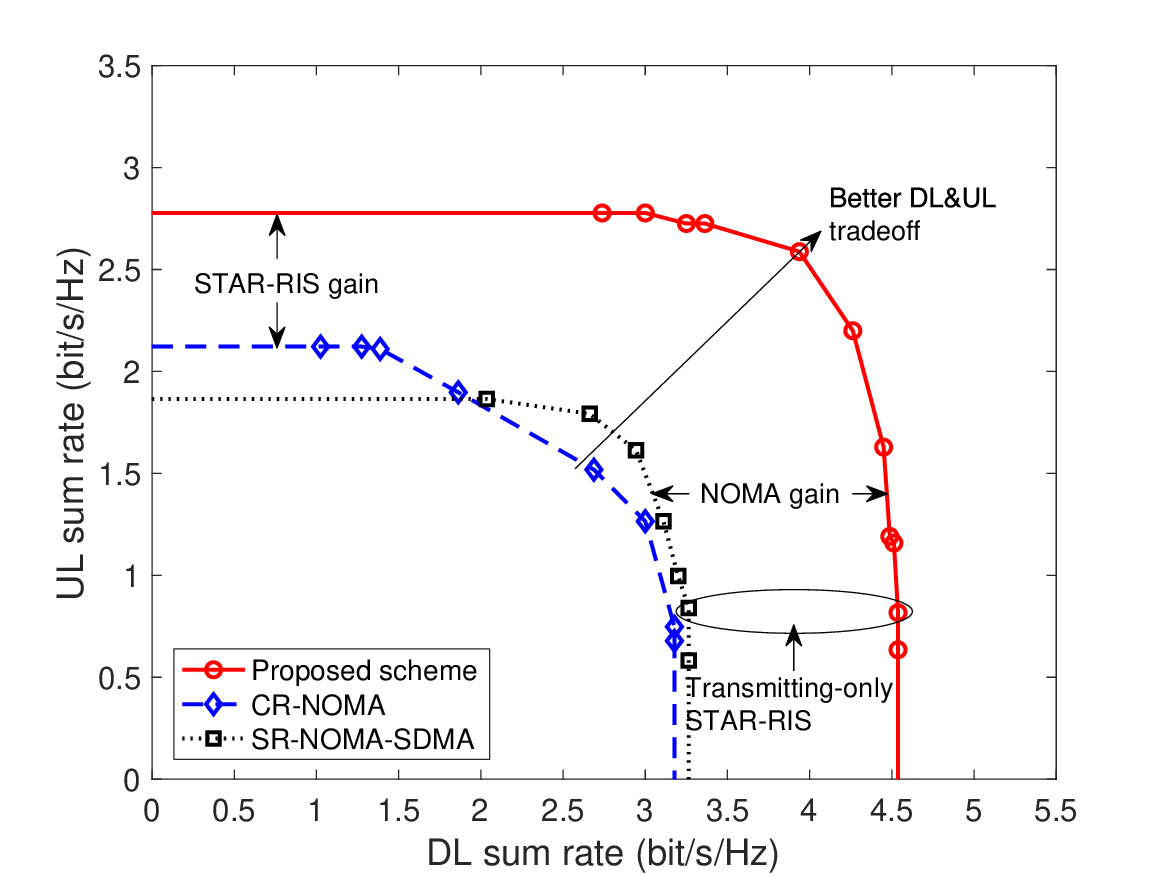}
	\caption{DL and UL transmission sum rate region. }
	\label{rate tradeoff}
\end{figure}
Fig. 6 characterizes the performance tradeoff between the DL and UL transmissions. With fixed $\omega^{\rm DL}=1$, by varying $\omega^{\rm UL}$ we can obtain the sum rate regions for both transmissions, respectively. It can be observed that the rate region achieved by the proposed scheme is significantly larger than that with other two baseline schemes, which underscores the performance advantage of our design. 
Furthermore, our proposed scheme and the SR-NOMA-SDMA scheme can obtain an ideal upper bound on the achievable rate of DL transmission, in which case the STAR-RIS employs a transmitting-only mode. As illustrated in the figure, the maximum UL sum rate is sometimes obtained at a cost of DL performance degradation, and an increase in the DL sum rate results in a loss of UL performance.  Compared with the CR-NOMA scheme, by employing the STAR-RIS, we can achieve a more flexible performance tradeoff.  
\subsection{Impact of the STAR-RIS Location}
\begin{figure}[t]
	\centering
	\setlength{\belowcaptionskip}{+0.2cm}   
	\includegraphics[width=3.3in]{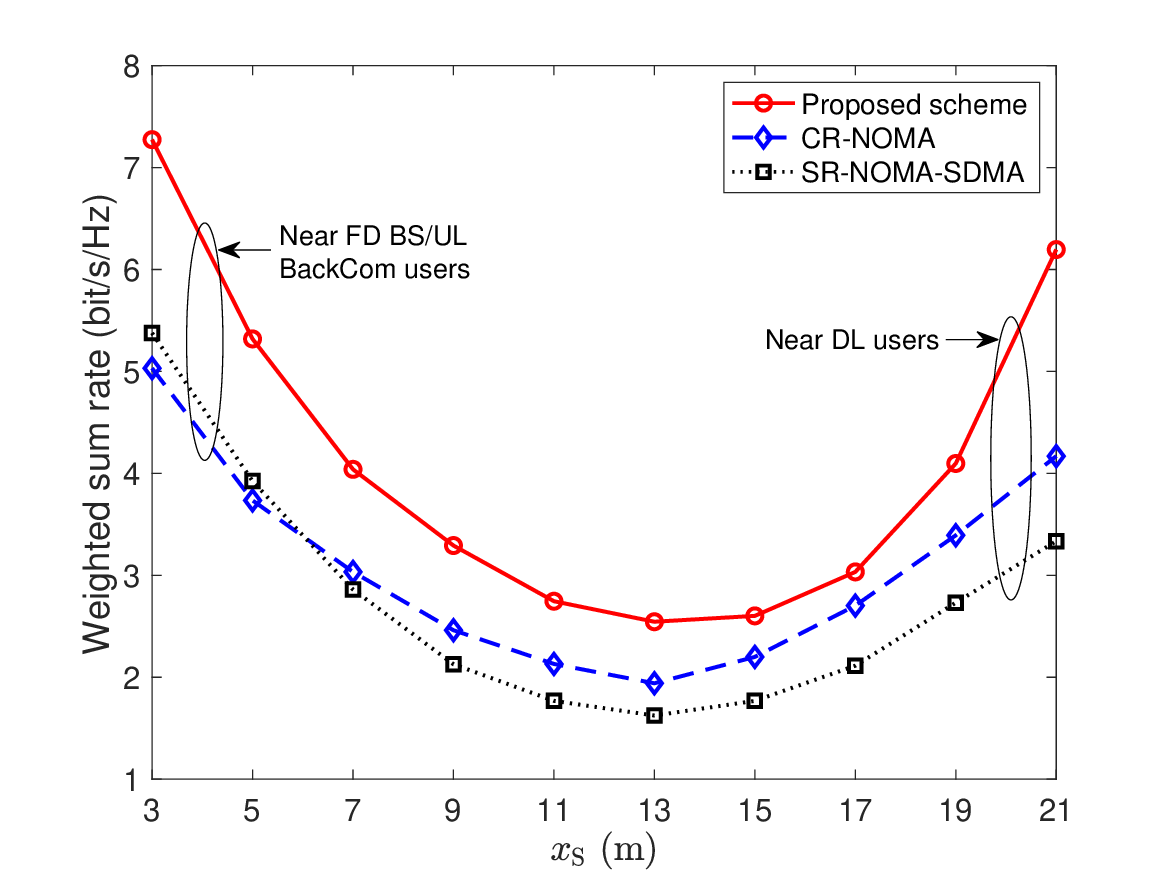}
	\caption{System weighted sum rate versus the location of STAR-RIS. }
	\label{STAR-RIS location}
\end{figure}
In Fig. 7, we depict the system weighted sum rate versus the location of STAR-RIS. The predefined coordinate of the STAR-RIS is set to be $(x_{\rm S}, \textup{0\;m, 0\;m})$, where $3\textup{\;m} \le x_{\rm S}\le 20\textup{\;m}$. It can be obsearved that, compared with the other two baseline schemes, our proposed scheme has significant performance improvement. For all the considered schemes, the system performance is highly sensitive to the deployment location of STAR-RIS. One more interesting finding is that, as is the pattern found in conventional RIS-assisted systems, higher performance gains can be achieved when STAR-RIS is closer to the FD BS or the users being served. These observations thus provide important guidelines for our practical implementation.
\subsection{Impact of the Number of Users}
\begin{figure}[t]
	\centering
	\setlength{\belowcaptionskip}{+0.2cm}   
	\includegraphics[width=3.3in]{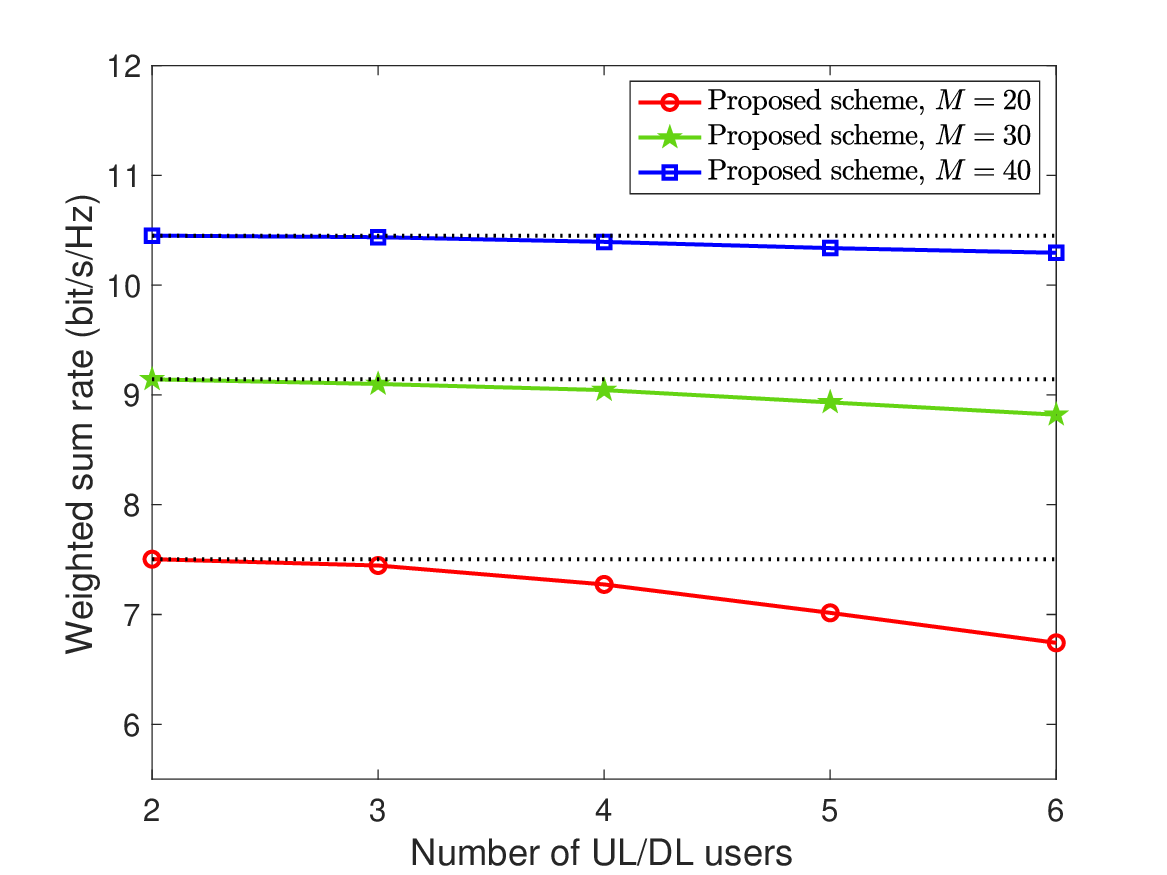}
	\caption{System weighted sum rate versus the number of users. }
	\label{user number}
\end{figure}
In Fig. 8, we investigate the system weighted sum rate versus the user number $I_{user}$ with different number of STAR-RIS elements. 
Note that the case of $M=40$ achieves the best performance, and while we can neglect the impact of the number of users on the system weighted sum rate, it also indicates that a larger number of users leads to a lower average rate for each individual user. Whereas, for the two baseline schemes, serving more users with fixed wireless resources causes a significant performance loss. From the above, we can conclude that as the increase in $M$ introduces higher array gains, deploying more elements on STAR-RIS is more robust against the tighter communication requirements.
\subsection{Impact of the DL NOMA Decoding Orders}
\begin{figure}[t]
	\centering
	\setlength{\belowcaptionskip}{+0.2cm}
	\includegraphics[width=3.3in]{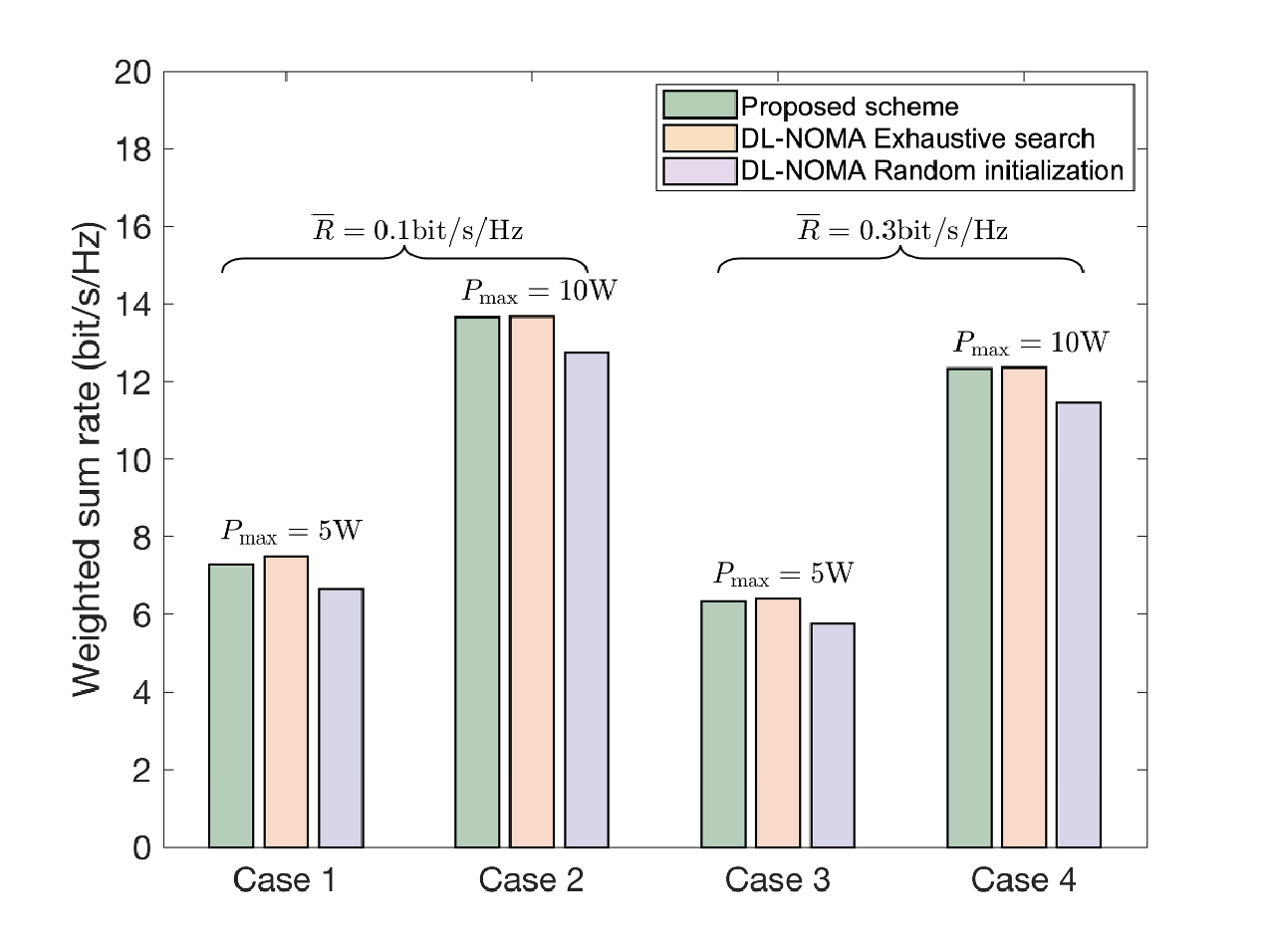}
	\caption{System weighted sum rate versus different DL NOMA decoding orders. }
	\label{decoding orders}
\end{figure}
To draw more useful insights, in Fig. 9, we provide the weighted sum rate versus the DL NOMA decoding orders. We consider four cases: (1) $P_{\max}=5$ W, $\overline R=0.1$ bit/s/Hz; (2) $P_{\max}=10$ W, $\overline R=0.1$ bit/s/Hz; (3) $P_{\max}=5$ W, $\overline R=0.3$ bit/s/Hz; and (4) $P_{\max}=10$ W, $\overline R=0.3$ bit/s/Hz. To evaluate our purposed scheme for initialing the DL decoding orders, two comparison options are considered, i.e., exhaustive search scheme and random initialization scheme. For the exhaustive search, we traversed all possible decoding orders and used the optimal solution as an upper bound on performance. For the random initialization scheme, we predefined a random decoding order for each DL user. As expected, the rate performance of the system deteriorates when the decoding order is random, which emphasises the need to optimize the NOMA decoding order. In considered multiuser scenario, the proposed initialization scheme is able to achieve near-optimal performance with relatively low-complexity, which is a favored strategy for NOMA DL decoding.
\section{Conclusions}
In this paper, a novel STAR-RIS assisted DL active and UL passive BackCom framework has been investigated, in which the NOMA protocol is exploited to fully reap the potential performance benefits of the hybrid transmission network. Based on the proposed setup, a joint active and passive beamforming optimization problem was formulated for maximizing the system weighted sum rate. To tackle this resulting non-trivial problem, an efficient algorithm is developed by invoking AO.
We further proposed a combined channel gain based user ordering strategy that achieves a comparable performance to the exhaustive search method, while entailing lower computational complexity. Simulation results verified the superiority of the proposed design over other baseline schemes. It revealed that, it is appealing to deploying STAR-RIS in a hybrid network, to enable simultaneous energy and spectrum cooperation between the active and passive transmissions. The obtained results also revealed that, the UL sum rate improvement is obtained at a cost of DL performance degradation. This insight provides a useful guide to tradeoff the performance between DL and UL transmissions for practical hybrid network implemention.

The results of this paper have confirmed the effectiveness of exploiting STAR-RISs in hybrid transmissions, which drives our follow-up research work. In particular, how to extend the network design to the system assisted by an STAR-RIS operating in time switching (TS) mode~\cite{mu2021simultaneously} is a promising direction worthy of future investigation. Unlike systems assisted by the STAR-RIS based on ES protocol, for which FD strategy could be unapplicable for transmissions, in addition, efficient channel estimation methods with tolerable pilot-overhead should be developed.
\section*{Appendix~A: Proof of Theorem~\ref{rank one relax}} \label{Appendix:A}
It is easy to prove the relaxed version of problem \eqref{P13} ignoring rank-one constraint \eqref{P12_C4} is jointly convex with respect to $\left\{\mathbf{W}_j,\mathbf{U}_t\right\}$ and the Slater's constraint qualification is satisfied~\cite{boyd2004convex}. Thus, strong duality holds and the derived Lagrangian function (60) is given at the top of the next page.
\begin{figure*}[!t]
	\normalsize
	\begin{equation}
		\begin{aligned} \label{eqn2}
			{{\mathcal{L}}} =&-\frac{1}{4}\sum_{j\in\mathcal{J}}\left\{2{\textup{Tr}}\left(\left({{{\mathbf U_t+{\mathbf{Q}}_j\mathbf{W}_j{\mathbf{Q}}_j^H}}} \right)\left({{{\widetilde{\mathbf U}_t+{{\mathbf{Q}}_j\widetilde{\mathbf{W}}_j{\mathbf{Q}}_j^H}}}}\right)^H\right)-\left\|{{{\mathbf U_t-{\mathbf{Q}}_j\mathbf{W}_j{\mathbf{Q}}_j^H}}} \right\|_F^2\right\}\\
			&+\lambda\sum_{j\in\mathcal{J}} {{\textup{Tr}}\left({{\mathbf{W}}_j} \right)} - \sum_{j\in\mathcal{J}} {{\textup{Tr}}\left( {{{\mathbf{Y}}_j}{{\mathbf{W}}_j}} \right)}+\varrho, 
		\end{aligned}
	\end{equation}
	\hrulefill \vspace*{0pt}
\end{figure*}
Here, $\varrho$ is introduced to collect all terms unrelated to $\left\{\mathbf{W}_j\right\}$, $\lambda$ and ${\mathbf{Y}}_j$ are the Lagrange multipliers correspondingly. Based on the Karush-Kuhn-Tucker (KKT) conditions, we provide the structure of the optimal ${\mathbf{W}}_j^*$ below:
\begin{subequations}\label{KKT}
	\begin{align}
		\label{KKT1}&\lambda^* \ge 0,{\mathbf{Y}}_j^* \succeq 0,\\
		\label{KKT2}&{\mathbf{Y}}_j^*{\mathbf{W}}_j^* = {\mathbf{0}},\\
		\label{KKT3}&{\nabla _{{\mathbf{W}}_j^*}}{{\mathcal{L}}} = 0,
	\end{align}
\end{subequations}
where $\lambda^*$ and ${\mathbf{Y}}_j^*$ denote the optimal Lagrange multipliers and ${\nabla _{{\mathbf{W}}_j^*}}{\mathcal{L}}$ is the gradient of ${\mathcal{L}}$ with respect to ${\mathbf{W}}_j^*$. In fact, (61c) can be equated to
\begin{align}\label{KKT=0}
	{\mathbf{Y}}_j^* =\lambda^*{{\mathbf{I}}_N}  - {\mathbf{\Lambda}}_j^*,
\end{align}
where ${\mathbf{\Lambda }}_j^*$ is given by
\begin{align}\label{Gamma}
	\begin{gathered}
		{\mathbf{\Lambda }}_j^* = \frac{1}{2}\left( {{{\mathbf{Q}}_j^H}{{\mathbf{F}}^H}{\mathbf{Q}}_j - {{\mathbf{Q}}_j^H}{{\mathbf{U}}_t}{{\mathbf{Q}}_j}+ {{\mathbf{Q}}_j^H}\mathbf{Q}}_j{{\mathbf{W}}_j^H}{{\mathbf{Q}}_j^H}{{\mathbf{Q}}_j}\right),
	\end{gathered}
\end{align}
where ${\mathbf{F}}\triangleq \left({{{\widetilde{\mathbf U}_t+{{\mathbf{Q}}_j\widetilde{\mathbf{W}}_j{\mathbf{Q}}_j^H}}}}\right)^H\in\mathbb{C}^{M \times M}$.
Referring to the results obtained in~\cite{xu2020resource}, it can be shown that ${\textup{Rank}}\left( {{\mathbf{Y}}_j^*} \right) = N - 1$. On the other hand, (61b) indicates that ${\textup{Rank}}\left( {{\mathbf{Y}}_j^*} \right) + {\textup{Rank}}\left( {{\mathbf{W}}_j^*} \right) \le N$. Therefore, ${\textup{Rank}}\left( {{\mathbf{W}}_j^*} \right) \le 1$ has to hold. Then, we can assert that the obtained optimal solution always satisfies ${\textup{Rank}}\left( {{\mathbf{W}}_j^*} \right) = 1$. In the light of the above conclusions, the proof is complete.

\balance
\bibliographystyle{IEEEtran}
\balance
\bibliography{mybib}

 \end{document}